\documentclass[twocolumn,prd,noshowpacs,nofootinbib,amsmath,amssymb]{revtex4}
\usepackage{graphicx,epsfig,psfrag,bm,amssymb}
\usepackage{dcolumn}
\usepackage{bm}
\usepackage{color}
\usepackage{mathrsfs,amsfonts,hepunits, color}
\usepackage{mciteplus}
\usepackage{tikz}
\usetikzlibrary{arrows,shapes}
\usetikzlibrary{trees}
\usetikzlibrary{matrix,arrows} 				
\usetikzlibrary{positioning}				
\usetikzlibrary{calc,through}				
\usetikzlibrary{decorations.pathreplacing}  
\usepackage{pgffor}							

\usetikzlibrary{decorations.pathmorphing}	
\usetikzlibrary{decorations.markings}
\usetikzlibrary{snakes}

\tikzset{
    vector/.style={decorate, decoration={snake}, draw},
	provector/.style={decorate, decoration={snake,amplitude=2.5pt}, draw},
	antivector/.style={decorate, decoration={snake,amplitude=-2.5pt}, draw},
    fermion/.style={draw, postaction={decorate},
        decoration={markings,mark=at position .55 with {\arrow[draw]{>}}}},
    fermionbar/.style={draw, postaction={decorate},
        decoration={markings,mark=at position .55 with {\arrow[draw=black]{<}}}},
    fermionnoarrow/.style={draw},
    gluon/.style={decorate, draw,decoration={coil,amplitude=4pt, segment length=6pt}, line width=1},
    scalar/.style={dashed,draw, postaction={decorate},
        decoration={markings,mark=at position .55 with {\arrow[draw]{>}}}},
    scalarbar/.style={dashed,draw, postaction={decorate},
        decoration={markings,mark=at position .55 with {\arrow[draw]{<}}}},
    scalarnoarrow/.style={dash pattern = on 6 pt off 3 pt,draw},
    electron/.style={draw, postaction={decorate},
        decoration={markings,mark=at position .55 with {\arrow[draw]{>}}}},
	bigvector/.style={decorate, decoration={snake,amplitude=4pt}, draw},
	vectorscalar/.style={loosely dotted,draw, postaction={decorate}},
}
\RequirePackage{xspace}
\usepackage{relsize}
\usepackage{slashed}
\newcommand{\fb}{{\rm fb}}
\newcommand{\ab}{{\rm ab}}

\newcommand{\be}{\begin{eqnarray}}
\newcommand{\ee}{\end{eqnarray}}
\def\lsim{\mathrel{\rlap{\lower4pt\hbox{\hskip 0.5 pt$\sim$}}
    \raise1pt\hbox{$<$}}}                
\def\gsim{\mathrel{\rlap{\lower4pt\hbox{\hskip1pt$\sim$}}
    \raise1pt\hbox{$>$}}} 
    
\newcommand{\schi}{s_{\chi\bar\chi}}  

\newcommand{\f}{\frac}
\newcommand{\pf}[2]{\left(\frac{#1}{#2}\right)}
\newcommand{\g}{{\rm g}}
\newcommand{\s}{{\rm s}}
\newcommand{\m}{{\rm m}}

\begin{document}

\title{ New Electron Beam-Dump Experiments to Search for MeV to few-GeV Dark Matter}  

\author{Eder Izaguirre, Gordan Krnjaic, Philip Schuster, and Natalia Toro}


\vspace{2cm}
\affiliation{ \\ Perimeter Institute for Theoretical Physics, Waterloo, Ontario, Canada                  }
\vspace{5cm}
\date{\today}

\begin{abstract}
In a broad class of consistent models, MeV to few-GeV dark matter interacts with ordinary matter through weakly coupled GeV-scale mediators.  
We show that a suitable meter-scale (or smaller) detector situated downstream of an electron beam-dump can sensitively probe dark matter interacting via sub-GeV mediators, while B-factory searches cover the 1--5 GeV range. 
Combined, such experiments explore a well-motivated and otherwise inaccessible region of dark matter parameter space with sensitivity 
several orders of magnitude beyond existing direct detection constraints. These experiments would also probe invisibly decaying new gauge bosons (``dark photons") down to kinetic mixing of $\epsilon\sim 10^{-4}$, including the range of parameters relevant for explaining the $(g-2)_{\mu}$ discrepancy. Sensitivity to other long-lived dark sector states and to new milli-charge particles would also be improved.
\end{abstract}

\maketitle

\section{Introduction and Summary}

\begin{figure}[t]
\includegraphics[width=8.6cm]{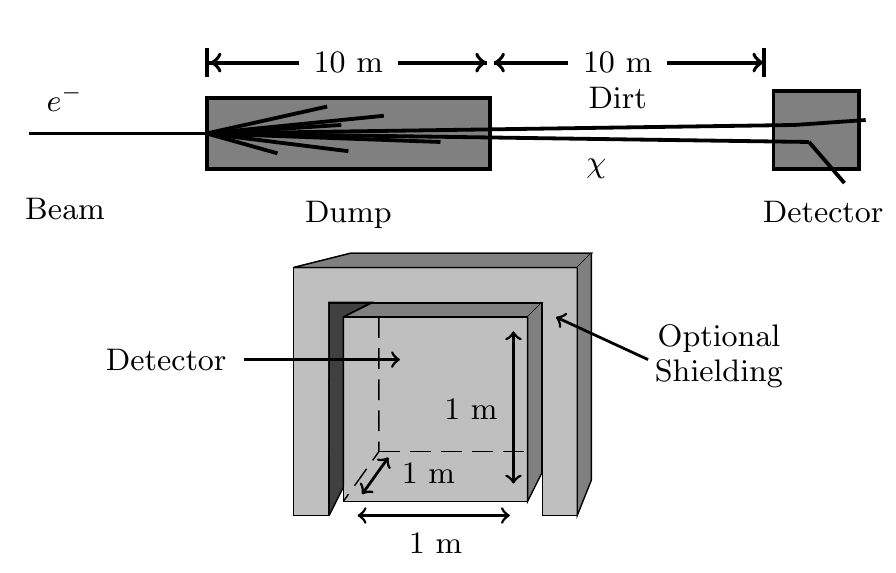}
\caption{Schematic experimental setup. A high-intensity multi-GeV electron beam impinging on a beam dump produces a secondary beam of dark sector 
states.  In the basic setup, a small detector is placed downstream so that muons and energetic neutrons are entirely ranged out. 
In the concrete example we consider, a scintillator detector is used to study quasi-elastic $\chi$-nucleon scattering at momentum transfers $\gsim 140$ MeV, 
well above radiological backgrounds, fast neutrons, and noise. 
Similar layouts with much smaller detectors or shorter target-detector distances than shown above are similarly sensitive. 
To improve sensitivity, additional shielding or vetoes can be used to actively reduce high energy cosmogenic
 and other environmental backgrounds. }\label{fig:setup}
\end{figure}

Dark matter is sharp evidence for physics beyond the Standard Model, and may be our first glimpse at a rich sector of new phenomena at accessible mass scales.
Whereas vast experimental programs aim to detect or produce few-GeV-to-TeV dark matter \cite{Aalseth:2012if,Aprile:2013doa,Ackermann:2011wa,Fermi-LAT:2013uma,Angloher:2011uu,Bernabei:2010mq,Agnese:2013rvf,Ahmed:2010wy,Chatrchyan:2012me,ATLAS:2012zim,Ackermann:2012qk,Aprile:2012nq}, these experiments are essentially blind to dark matter of MeV-to-GeV mass.  
We propose an approach to search for dark matter in this lower mass range by producing it in an electron beam-dump and then detecting its scattering in a small downstream detector (Fig.~\ref{fig:setup}).  
This approach can explore significant new parameter space for both dark matter and light force-carriers decaying invisibly,  in parasitic low-beam-background experiments at existing facilities.   The sensitivity of this approach complements and extends that of analogous proposed neutrino factory searches \cite{Batell:2009di,deNiverville:2011it,deNiverville:2012ij,deNiverville:2013}. 
Combined with potential B-factory searches, these experiments would explore a well-motivated and otherwise inaccessible region of dark matter parameter space.
Experiments of this type are also essential to a robust program searching for new kinetically mixed gauge bosons, as they complement the ongoing searches for such bosons' visible decays \cite{Bjorken:2009mm,Batell:2009di,deNiverville:2011it,Bjorken:1988as,Riordan:1987aw,Bross:1989mp,Essig:2009nc,Essig:2010xa,Batell:2009yf,Fayet:2007ua,Freytsis:2009bh,Essig:2010gu,Reece:2009un,Wojtsekhowski:2009vz,AmelinoCamelia:2010me,Baumgart:2009tn,Merkel:2011ze,Abrahamyan:2011gv,Aubert:2009cp,Babusci:2012cr,Echenard:2012hq,Andreas:2012mt,Adlarson:2013eza}.

Various considerations motivate dark matter candidates in the MeV-to-TeV range.
Much heavier dark matter is disfavored because its naive thermal abundance exceeds the observed cosmological matter density.
Much beneath an $\MeV$, astrophysical and cosmological constraints allow only dark matter with ultra-weak couplings to quarks and leptons \cite{Hewett:2012ns}.  Between these boundaries ($\MeV - \TeV$), simple models of dark matter can account for its observed abundance through either thermal freeze-out or non-thermal mechanisms \cite{Kaplan:1991ah,Barr:1990ca,ArkaniHamed:2008qn,Hooper:2008im,Feng:2008ya,Feng:2008mu,Feng:2010tg,Pospelov:2008jd,Katz:2009qq,Kaplan:2009ag,Kaplan:2011yj,Falkowski:2011xh,Alves:2009nf,Hooper:2012cw,Cline:2012is,CyrRacine:2012fz}.  The conventional argument in favor of weak-scale  ($\gtrsim 100\ \GeV$) dark matter --- that its annihilation through Standard Model (SM) forces alone suffices to explain the observed relic density --- is dampened by strong experimental constraints on dark matter with significant couplings to the $Z$ or Higgs bosons \cite{Aprile:2012nq,Cirelli:2009uv} and by the absence to date of evidence for new SM-charged matter at the LHC.  

The best constraints on multi-GeV dark matter interactions are from underground searches for nuclei recoiling off non-relativistic dark matter particles in the Galactic halo (e.g. \cite{Aalseth:2012if,Aprile:2013doa,Aprile:2012nq,Angloher:2011uu,Bernabei:2010mq,Agnese:2013rvf,Ahmed:2010wy,Chatrchyan:2012me}).  These searches are insensitive to few-GeV or lighter dark matter, whose nuclear scattering transfers  invisibly small kinetic energy to a recoiling nucleus.  Electron-scattering offers an alternative strategy to search for sub-GeV dark matter, but with dramatically higher backgrounds \cite{Essig:2012yx,Essig:2011nj,Graham:2012su}.
If dark matter scatters by exchange of particles \emph{heavier} than the $Z$, then competitive limits can be obtained from hadron collider searches for dark matter pair-production accompanied by a jet, which results in a high-missing-energy ``monojet'' signature \cite{Chatrchyan:2012me,ATLAS:2012zim}.  But among the best motivated models of $\MeV-\GeV$ dark matter are those whose interactions with ordinary matter are mediated by new $\GeV$-scale ``dark'' force carriers (for example, a gauge boson that kinetically mixes with the photon) \cite{Pospelov:2007mp,ArkaniHamed:2008qn}.  Such models readily account for the stability of dark matter and its observed relic density, are compatible with observations, and have important implications beyond the dark matter itself.  In these scenarios, high energy accelerator probes of sub-GeV dark matter are as ineffective as direct detection searches, because the missing energy in dark matter pair production is peaked well below the $Z\rightarrow \nu\bar\nu$ background and is invisible over QCD backgrounds\cite{An:2012va,Fox:2011pm}.

Instead, the tightest constraints on light dark matter arise from B-factory searches in (partly) invisible decay modes \cite{Aubert:2008as}, rare kaon decays \cite{Artamonov:2008qb}, precision $(g-2)$ measurements of the electron and muon \cite{Pospelov:2008zw,Giudice:2012ms}, neutrino experiments \cite{deNiverville:2013}, supernova cooling, and high-background analyses of electron 
recoils in direct detection \cite{Essig:2012yx}. 
These constraints and those from future B-factories and neutrino experiments leave a broad and well-motivated 
class of sub-GeV dark matter models largely unexplored.
For example, with a dark matter mass $\gsim 70 \ \MeV$, existing neutrino factories and optimistic projections for future 
Belle II sensitivity leave a swath of parameter space relevant for reconciling the $(g-2)_{\mu}$ anomaly wide open (see Figure \ref{fig:SummaryReach}). 
More broadly, the interaction strength best motivated in the context of models with kinetically mixed force carriers (mixing $10^{-5}\lesssim \epsilon \lesssim 10^{-3}$) lies just beyond current sensitivity across a wide range of dark matter and force carrier masses in the $\MeV-\GeV$ range.  
These considerations, along with the goal of greatly extending sensitivity to {\it any} components of $\MeV-\GeV$ dark matter 
beyond direct detection constraints motivates a much more aggressive program of searches in the coming decade. 

\begin{figure}[ht]
\includegraphics[width=8.6cm]{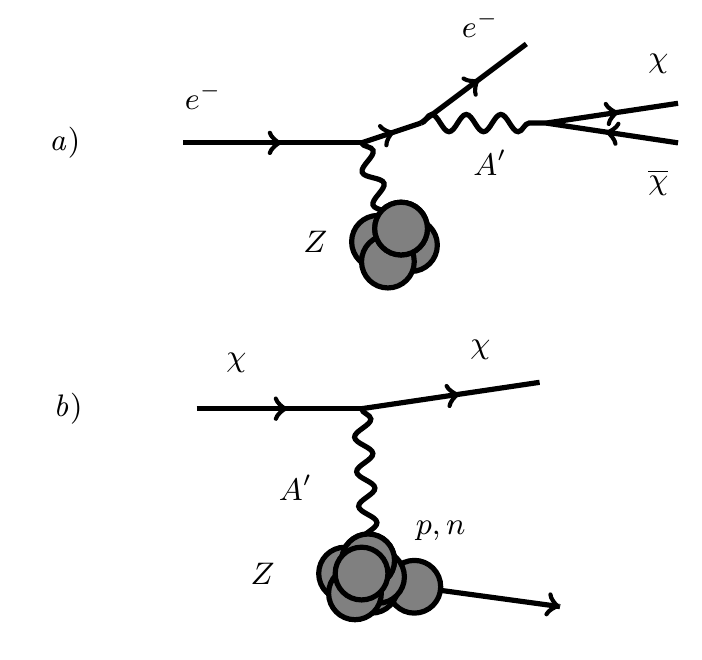}
           \caption{  a) $\chi \bar \chi$ pair production in electron-nucleus collisions via the Cabibbo-Parisi radiative process (with $A'$ on- or off-shell) and b) 
$\chi$  scattering off a detector nucleus and liberating a constituent nucleon. 
For the momentum transfers of interest, the incoming $\chi$ resolves the nuclear substructure, so the typical reaction is quasi-elastic and 
nucleons will be ejected.
 }\label{fig:prod}
\end{figure}

\begin{figure*}
 \vspace{0.1cm}
\includegraphics[width=8.6cm]{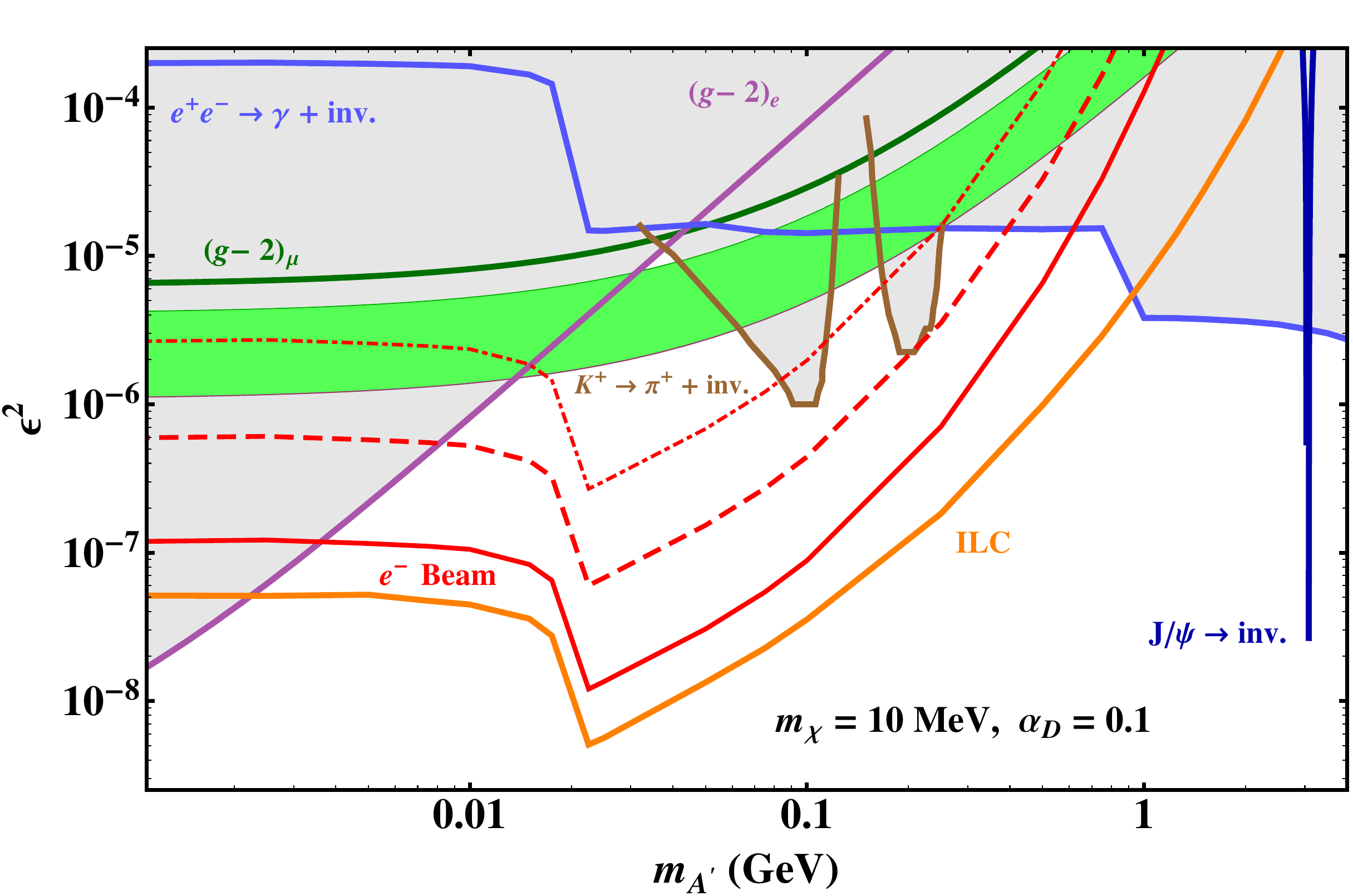}
\includegraphics[width=8.6cm]{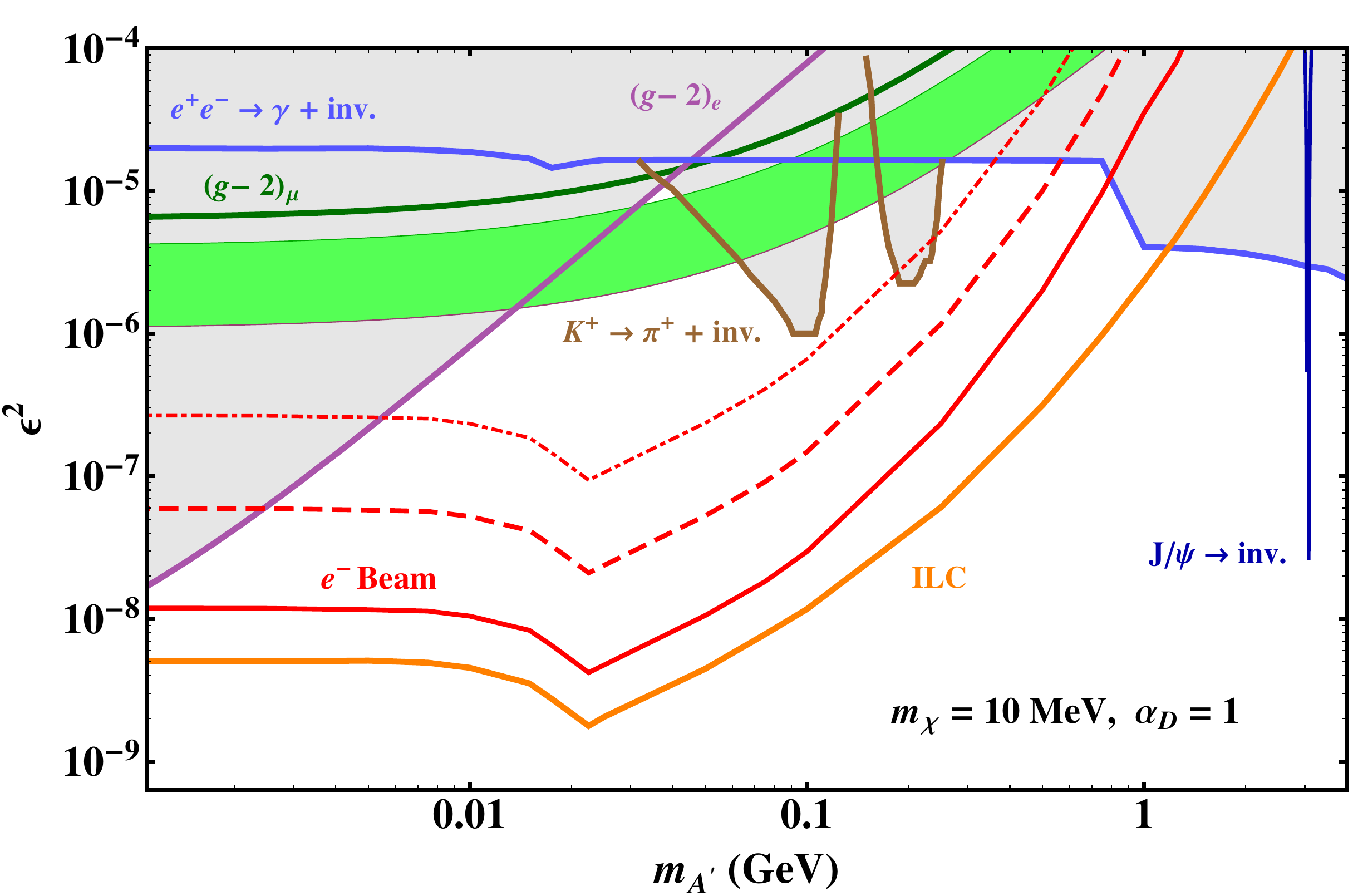} 
\includegraphics[width=8.6cm]{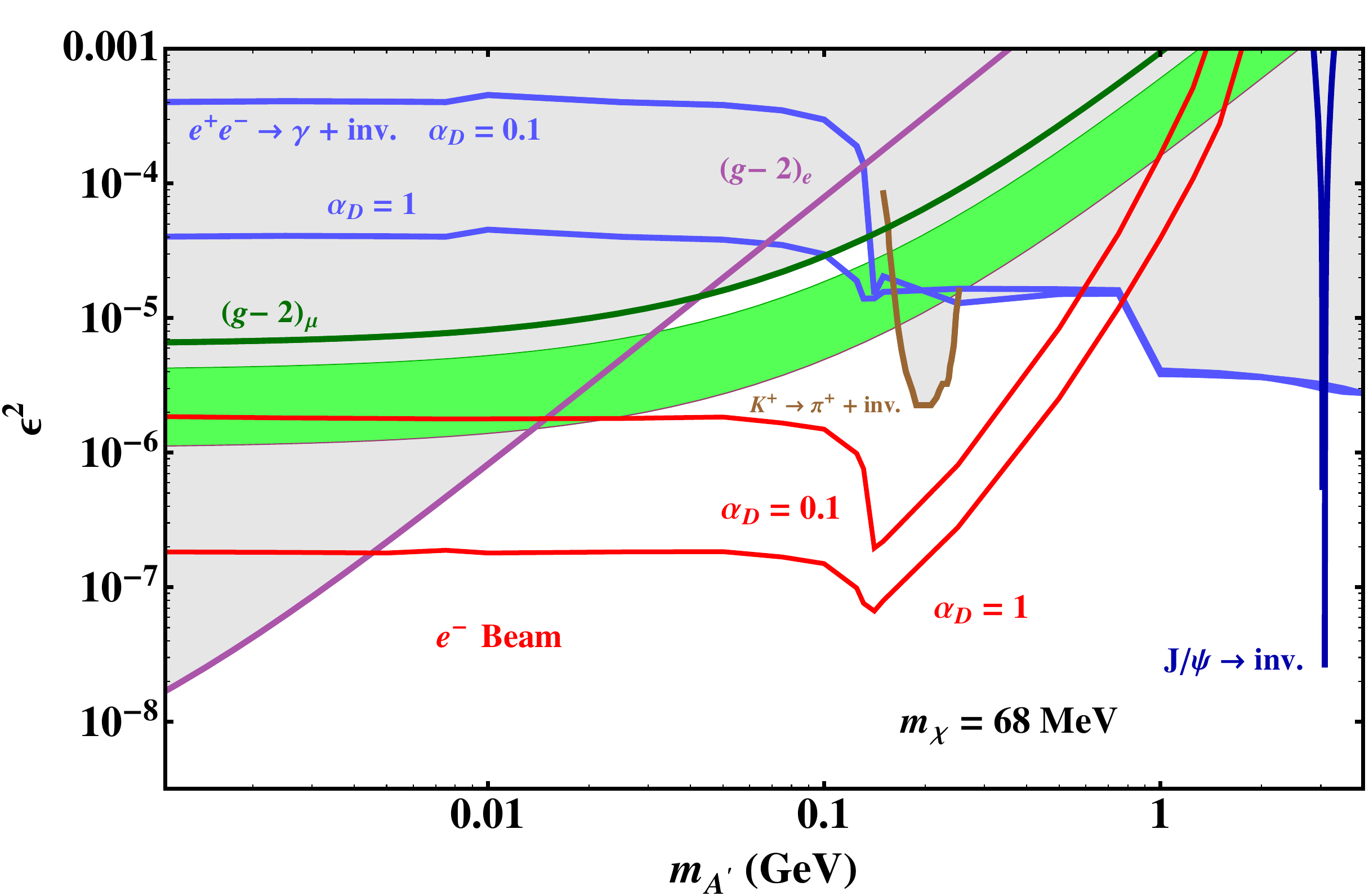} 
\includegraphics[width=8.6cm]{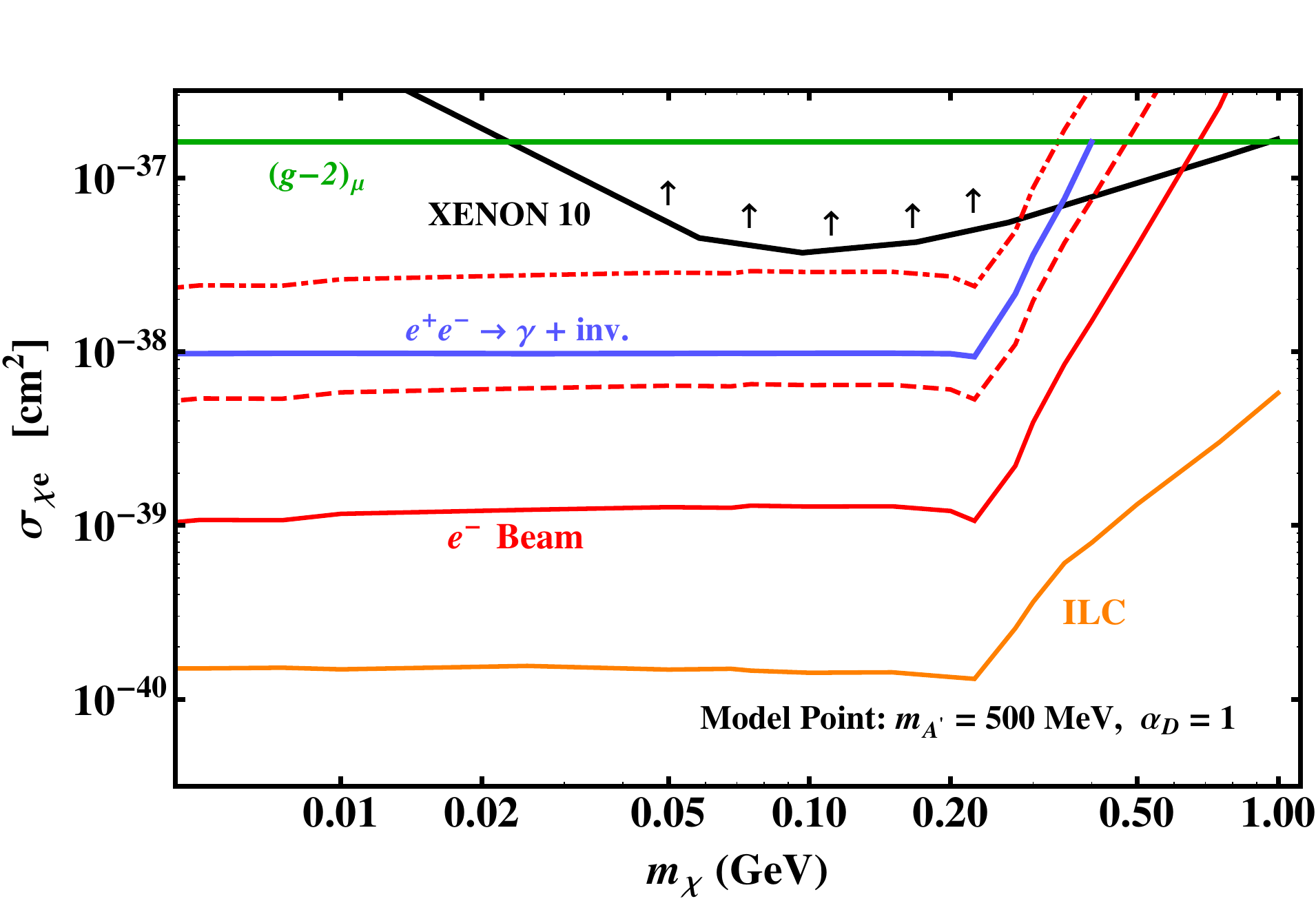}
  \caption{The $\epsilon^2$ sensitivity of electron-beam 
  fixed-target experiments plotted alongside existing constraints for benchmark values of $m_\chi$, $m_{A'}$, and $\alpha_D$. 
   The solid, dashed, and dot-dashed red curves mark the parameter space for which our basic setup --- a $12~\GeV$ beam impinging on an aluminum beam dump, with a 1\,m$^3$ mineral oil detector placed 20\,m downstream of the dump --- 
   respectively yields  40, $10^3$, and $2 \cdot 10^4$ $\chi$-nucleon quasi-elastic scattering events with $Q^2 \gtrsim (140~\MeV)^2$ per $10^{22}$ electrons on target (EOT). The orange
curve shows the 10 event reach for an ILC style  
 125 GeV beam assuming the same detector and luminosity.  
 Comparable sensitivity can be achieved with much smaller fiducial volumes than we consider, especially for detectors with active 
 muon and neutron shielding and/or veto capabilities. The upper plots show the $\epsilon$ sensitivity for $\alpha_D =0.1$ (left) and $\alpha_D =1$ (right).
In these plots LSND may also have sensitivity to $\epsilon^2 \sim 10^{-8} -10^{-6}$ via $\pi^0 \to \gamma \chi \bar \chi$ decays for
     $ 2 m_{\chi} < m_{A'} < m_{\pi}$ \cite{deNiverville:2013}. The lower left plot shows the reach for $m_\chi = m_{\pi^0} /2 \simeq 68$ MeV where the  production from
     pion decays is  kinematically inaccessible and LSND has no significant sensitivity. 
The lower right plot recasts the $\epsilon^2$ sensitivity for fixed $m_{A'}$ and $\alpha_D$ as a (model-dependent) probe of the $\chi$-electron direct detection cross
      section $\sigma_{\chi e} $ and includes XENON 10 limits from \cite{Essig:2012yx}. The black curve assumes $\Omega_\chi = \Omega_{DM}$; the 
      direct detection constraint is weaker when $\chi$ is only a component of the total abundance.  
      The light green band is the region in which an $A^\prime$ resolves the $(g-2)_\mu$ discrepancy to within $2 \sigma$; the 
   dark green curve is the  boundary at which contributions to $(g-2)_\mu$ exceed the
    measured value by $5 \sigma$ \cite{Pospelov:2008zw}. 
    The bound from $e^+ e^- \to \gamma \, +$ invisibles is introduced in detail in section \ref{ssec:collider}
  Other constraints in the literature arise from invisible $J/\psi$ decays \cite{Ablikim:2007ek}
   searches \cite{Aubert:2008as}, rare kaon decays \cite{Artamonov:2008qb}, and contributions to $(g-2)_e$  \cite{Giudice:2012ms}; for a 
   discussion see section \ref{ssec:otherlab}.
   }\label{fig:SummaryReach}
\end{figure*}

The experimental setup we consider can dramatically extend sensitivity to long-lived weakly coupled states 
(see Fig.~\ref{fig:SummaryReach}), including $\GeV$-scale dark matter, any \emph{component} of dark matter below a few GeV, 
and milli-charged particles. This includes a swath of light force carrier parameters motivated by the $(g-2)_{\mu}$ anomaly, 
extending beyond the reach of proposed neutrino-factory searches and Belle II projections (see Figure \ref{fig:SummaryReach}).
The setup requires a small $1\ \m^3$-scale (or smaller) detector volume tens of meters downstream of the 
beam dump for a high-intensity multi-GeV electron beam (for example, behind the Jefferson Lab Hall A or C dumps or a linear collider beam dump), 
and could run parasitically at existing facilities (see \cite{Diamond} for a proof-of-concept example).
All of the above-mentioned light particles (referred to hereafter as ``$\chi$'') can be pair-produced radiatively in electron-nucleus collisions 
in the dump (see Fig.~\ref{fig:prod}a).  A fraction of these relativistic particles then scatter off nucleons, nuclei, or electrons in the detector volume (see Fig.~\ref{fig:prod}b).

Within a year, Jefferson Laboratory's CEBAF (JLab) \cite{2011JPhCS.312c2014M} will produce $100 \mu A$ beams at 12 GeV.  
Even a simple meter-scale (or smaller) instrument capable of detecting quasi-elastic nucleon scattering, but without cosmic background rejection, 
positioned roughly 20 meters (or less) downstream of the Hall A dump has interesting physics sensitivity (upper, dotted red curves in Fig.~\ref{fig:SummaryReach}).  
Dramatic further gains can be obtained by shielding from or vetoing cosmogenic neutrons (lower two red curves), or more simply 
by using a pulsed beam.  The lower red curve corresponds to $40$-event sensitivity per $10^{22}$ electrons on target, 
which may be realistically achievable in under a beam-year at JLab.  
The middle and upper red curves correspond to background-systematics-limited configurations, with $1000$ and $2\cdot10^4$ signal-event sensitivity, respectively, per $10^{22}$ electrons on target. Though not considered in detail in this paper, detectors sensitive to $\chi$-electron elastic scattering, coherent $\chi$-nuclear scattering, and pion production in inelastic $\chi$-nucleon scattering could have additional sensitivity. 
 With a pulsed beam, comparable parameter space could be equally well probed with 1 to 3 orders of magnitude less intensity. A high-intensity pulsed beam such as the proposed ILC beam could reach even greater sensitivity (orange curve).  
The parameter spaces of these plots are explained in the forthcoming subsection. 

The beam dump approach outlined here is quite complementary to B-factory $\gamma+invisible$ searches \cite{Aubert:2008as}, 
with better sensitivity in the $\MeV - \GeV$ range and less sensitivity for $1-10 \ \GeV$ (see also \cite{LightDMCollider}).
Compared to similar search strategies using proton beam dumps, the setup we consider has several virtues.
Most significantly, beam-related neutrino backgrounds, which are the limiting factor for proton beam setups, are negligible for electron beams. 
MeV-to-GeV $\chi$ are also produced with very forward-peaked kinematics (enhanced at high beam energy), 
permitting large angular acceptance even for a small detector. 
Furthermore, the expected cosmogenic backgrounds are known, measurable \emph{in situ}, and systematically reducible; with a pulsed electron beam, 
beam timing alone dramatically reduces these backgrounds.

The plan of this paper is as follows. In the remainder of this introduction, we summarize the discovery potential of electron beam dump
experiments that can be readily carried out within the next few years, and highlight their complementarity with other searches for dark sector particles. 
 In Section \ref{Sec: Simple Models} we discuss several viable scenarios for MeV$-$GeV scale dark matter and present an explicit model.
In Section \ref{Sec: Constraints} we summarize existing constraints on light $\chi$ interacting with ordinary matter through 
kinetically mixed gauge bosons. 
Among these, the B-factory and supernova constraints discussed in parts \ref{ssec:collider} and \ref{ssec:supernova} have not been previously considered in the literature.
In Section \ref{Sec: ProductionDetection}, we discuss the production of long-lived dark sector states and their scattering in the detector, providing approximate formulae so that the reader can easily rescale our results to other geometries and beam energies.   
In Section \ref{Sec: ExperimentalApproach}, we first discuss expected beam-related and cosmogenic backgrounds for a benchmark scenario modeled on JLab CEBAF-12 parameters, in which a meter-scale detector sensitive to neutral-current scattering is situated 20 meters downstream of an aluminum beam-dump for a 12 GeV, 80 $\mu A$ electron beam.  We then estimate the sensitivity of such a detector in several background-rejection scenarios, and illustrate the impact on sensitivity of various changes to model and detector parameters.  In Section \ref{Sec:ProtonComparison} we compare our approach with existing proposals and searches 
for dark sector states at neutrino factories. Finally, in Section \ref{Sec:Conclusion} we offer some concluding remarks and suggest future studies to improve upon our 
projections. 

\subsection{Discovery Potential}
We focus for concreteness on scenarios where the dark matter is part of a ``dark sector'' with its own gauge interactions.
Theories with light force-mediators are well-motivated in the context of sub-GeV dark matter as they permit relatively efficient annihilation of the light dark matter, preventing its relic density from exceeding the observed dark matter density.  
Independently of light dark matter, such scenarios have received tremendous attention in recent years \cite{Holdom:1985ag,Strassler:2006im,Pospelov:2007mp,Pospelov:2007mp,ArkaniHamed:2008qn,ArkaniHamed:2008qp,Morrissey:2009ur,Chang:2010yk,Andreas:2011in,Hooper:2012cw}, offer novel explanations of dark matter
compatible with existing CMB and galactic observations  \cite{Pospelov:2007mp,Huh:2007zw,Pospelov:2008jd}, and are the target of a growing international program 
of searches  \cite{Bjorken:1988as,Riordan:1987aw,Bross:1989mp,Essig:2009nc,Essig:2010xa,Fayet:2007ua,Freytsis:2009bh,Batell:2009di,Essig:2010gu,Reece:2009un,Wojtsekhowski:2009vz,AmelinoCamelia:2010me,Batell:2009yf,Baumgart:2009tn,deNiverville:2011it,Merkel:2011ze,Abrahamyan:2011gv,Aubert:2009cp,Babusci:2012cr,Echenard:2012hq,Andreas:2012mt,Adlarson:2013eza}. 

A dark matter component $\chi$ produced in fixed-target collisions can be a fermion or scalar, its abundance can arise thermally or non-thermally and be matter-symmetric or asymmetric \cite{Kaplan:1991ah,Kaplan:2009ag, Steffen:2006hw, Hamaguchi:2009hy, Petraki:2013wwa,Lin:2011gj}, and can comprise the full or a sub-dominant fraction of the cosmological dark matter density.  
Indeed, a thermal relic abundance of GeV-scale dark matter consistent with observation arises most naturally in models with a dark sector worthy of the name that, like the Standard Model sector, contains multiple light particles and multiple gauge forces, as discussed in Section \ref{Sec: Simple Models}.
Even if the dark sector is quite complicated, the fixed-target phenomenology of stable $\chi$ (or unstable $\chi$ with lab-frame lifetimes $\gtrsim \mu\s$) is usually well-described by the simplest case of a $U(1)_D$ dark sector with a single stable matter particle $\chi$, e.g. for fermionic $\chi$ 
 \be
 \label{eq:lagrangian}
{\cal L}_{dark} &=& 
-\frac{1}{4}F^\prime_{\mu\nu} F^{\prime\,\mu\nu} + \frac{\epsilon_Y}{2} F^\prime_{\mu\nu} B_{\mu \nu} + \frac{m^2_{A^\prime}}{2} A^{\prime}_\mu A^{\prime\, \mu} \nonumber \\
&&+  \bar \chi ( i \displaystyle{\not}{D}- m_\chi) \chi,
 \ee
where $B_{\mu \nu} = B_{[\mu,\nu]}$ and  $F^\prime_{\mu \nu} =
 A^\prime_{[\mu,\nu]}$ are respectively the  hypercharge and dark-photon field strengths and $D_\mu  = \partial_\mu + i g_D A^\prime_\mu$ (and similarly for scalar $\chi$).
 The kinetic mixing parameter $\epsilon_Y$ can arise generically from loops of heavy particles charged under both hypercharge and $U(1)_D$
 and is naturally small, on the scale of $\frac{e g_D}{16\pi^2} \log(M/\Lambda) \sim 10^{-3}-10^{-2}$ if it arises from loops of a mass-$M$ particle in a theory with cutoff scale $\Lambda$, and suppressed by an additional Standard Model loop factor if hypercharge is embedded in a unified gauge group at high scales.  In this paper, we will take $\epsilon\equiv \epsilon_Y \cos\theta_W$ (where $\theta_W$ is the weak mixing angle) to be a free parameter varying from roughly $10^{-5}$ to $10^{-2}$. As is well known, upon diagonalizing the kinetic mixing terms in \eqref{eq:lagrangian}, ordinary electrically charged matter acquires a ``dark millicharge'' coupling to the $A^\prime$ of strength $\epsilon e$, while $\chi$ remains electrically neutral. 
Long-lived dark sector particles $\chi$ couple to ordinary matter primarily through $A'$ exchange. This model also encompasses millicharged $\chi$ by taking the limit $m_{A'} \rightarrow 0$, in which case $\epsilon$ plays the role of the millicharge of $\chi$ in units of $e$.  We focus in this paper on fermionic $\chi$, but the same approach is sensitive to scalar $\chi$ as well.  

If $m_{A^\prime} < 2 m_\chi$,  the dominant $\chi$ production mechanism in
an electron fixed-target experiment is the radiative process illustrated in 
Fig. \ref{fig:prod}a) with off-shell $A'$. In this regime, the $\chi$ production yield scales as $\sim \alpha_D \epsilon^2/m^2_\chi$, 
while $\chi$-nucleon scattering in the detector via $A^{\prime}$ exchange, depicted in Fig. \ref{fig:prod}b),
 occurs with a rate proportional to $ \alpha_D  \epsilon^2/m_{A'}^2$ over most of the plotted mass range. 
 Thus, the total signal yield scales as
\be
N_{\chi} \sim   \frac{  \alpha_D^2 \epsilon^4 }{m_{\chi}^2m_{A'}^2}  ~~
\ee
(with additional suppression at high masses from loss of acceptance and nuclear coherence).
If $m_{A^\prime} > 2 m_\chi$, the secondary $\chi$-beam arises from
 radiative $A^{\prime}$ production followed by $A^\prime \to \bar \chi \chi$ decay.  In this regime, the $\chi$ production and
  the detector scattering rates are respectively proportional to $\epsilon^2/m_{A'}^2$  and $  \alpha_D  \epsilon^2/m_{A'}^2$, so the signal yield scales as  
\be
N_{\chi} \sim  \frac{ \alpha_D \epsilon^4}{m_{A'}^4} ~~.
\ee
Thus, for each $\alpha_D$ and $m_{A'}$, we can extract an $\epsilon$-sensitivity corresponding to a given scattering yield.

The characteristic momentum transfer in $\chi$-matter interactions is of order the $A'$ mass.
Low-momentum-transfer $\chi$-nucleus scattering in the detector, relevant for the smallest $A'$ masses, yield several distinct signals depending on the target nuclei. 
For most materials, scattering with momentum transfers below $ \sim 10$ MeV features a coherent $Z^2$ enhancement in the elastic cross section; 
the recoiling nucleus typically generates copious phonons, scintillation light, and (if kinematically allowed) Cerenkov photons.
 However, the enhanced signal rate in this energy range competes with ubiquitous radiological backgrounds, while far
  above 10 MeV, this process suffers sharp form-factor suppression.  Sensitivity to coherent scattering is 
  an experimental challenge, which deserves a dedicated study. In this energy regime elastic $\chi$-electron scattering can also
  yield large signal rates and may dominate the signal yield depending on the material and the cuts. 
  
For momentum transfers above $\sim$ 10 MeV, incoming $\chi$ resolve nuclear substructure and 
there is a rich variety of $\chi$-nucleon scattering channels. The dominant process in 
  a Carbon-based detector is quasi-elastic $\chi p,n \to \chi p,n$ scattering where a nucleon is liberated from the 
  nucleus, but appreciable yields can also arise from 
  resonant single pion production via  $\chi p \to \chi p \pi^0$ and $\chi p(n) \to \chi n (p) \pi^{+(-)}$. For  
  momentum transfers above $\sim 1$ GeV, resonant 
   $\chi p(n) \to \chi \Delta^0 \pi^+, \chi \Delta^0 \pi^+, \chi \Delta^{++} \pi^-$ and nonresonant $ \chi p \to \chi p \pi^+ \pi^- $ double-pion production
    may also be important. In principle, all of these signals should be studied, but our analysis
   in this paper focuses on quasi-elastic scattering off nucleons in mineral-oil, for which efficiencies are well known 
   \cite{Perevalov:2009zz}. 

Figure~\ref{fig:SummaryReach} summarizes estimates of the reach of our approach
quantified by the sensitivity to $\epsilon$ as a function of $m_{A'}$, $\alpha_D$, and $m_{\chi}$. 
Note that even a test version of this experiment without any background rejection (dot-dashed red curve) can have appreciable  
reach extending sensitivity to light dark matter by orders of magnitude.  

To gain some intuition for the power of this setup, it is instructive to compare our proposal against  
direct detection efforts. If $\chi$  comprises all of the dark matter, the most sensitive direct detection probe 
the MeV $-$ GeV range uses electron scattering, for which the cross section is roughly  
\be
\sigma_{\chi e} \sim \alpha_D\epsilon^2\frac{m_e^2}{m_{A'}^4}~~.
\ee
Thus, for a given relic density, $\alpha_D$, and $m_{A'}$, the $\epsilon$-sensitivity of accelerator-based experiments can be reinterpreted as a probe of 
$\sigma_{\chi e}$, enabling a (rather model-dependent) comparison with direct detection. In Fig.~\ref{fig:SummaryReach} (bottom right)
 we translate the reach in $\epsilon$ into a $\sigma_{\chi e}$ sensitivity, plotted alongside
 the bound from XENON 10 \cite{Essig:2012yx} assuming $\Omega_\chi = \Omega_{DM}$. 
 We see that the fixed target approach exceeds existing direct detection sensitivity by 
  orders of magnitude in cross section, for interactions modeled by \eqref{eq:lagrangian}. Importantly, the fixed target approach is also sensitive to highly subdominant components 
  of sub-GeV dark matter (which one may argue is even a natural expectation for these models), and in this 
  case the bounds from XENON 10 are weakened and fall off the above plot.  

The beam dump approach outlined here is quite complementary to B-factory $\gamma+invisible$ searches, with better sensitivity in the $\MeV - \GeV$ range and less sensitivity for $1-10 \ \GeV$.  The light blue curves in Fig.~\ref{fig:SummaryReach} show that the constraints from the existing mono-photon search at BaBar in the $A'$ mass range from $1-10\ \GeV$ will not be easily surpassed by beam-dump searches.  Moreover, the BaBar search is statistics-limited in this mass range, so that a similar search at Belle II may improve sensitivity by an order of magnitude.  However, as discussed in Sec.~\ref{ssec:collider}, for $\MeV - \GeV$ mass $A'$ the $B$-factory searches are limited by an instrumental background that mimics the $A'$ signal, so dramatic improvements from increased luminosity are unlikely.  
The sensitivity of the beam dump approach to sub-GeV masses is therefore particularly important.

These experiments probe rather inclusively the set of models where a kinetically mixed gauge boson decays \emph{invisibly} into dark matter, a sub-dominant component of dark matter, or metastable dark-sector particles.   They are therefore complementary to the ongoing searches for  MeV-to-GeV-mass gauge bosons decaying \emph{visibly} (either directly to leptons \cite{Bjorken:1988as,Riordan:1987aw,Bross:1989mp,Essig:2009nc,Essig:2010xa,Fayet:2007ua,Freytsis:2009bh,Batell:2009di,Essig:2010gu,Reece:2009un,Wojtsekhowski:2009vz,AmelinoCamelia:2010me,Batell:2009yf,Baumgart:2009tn,deNiverville:2011it,Merkel:2011ze,Abrahamyan:2011gv,Aubert:2009cp,Babusci:2012cr,Echenard:2012hq,Andreas:2012mt,Adlarson:2013eza} or indirectly through prompt dark-sector cascades \cite{Essig:2009nc,Andreas:2012mt,AmelinoCamelia:2010me}), and remarkably comparable in coupling sensitivity.  
The combined program of searches for light gauge bosons will rather decisively test whether the photon kinetically mixes with a GeV-scale gauge boson.  In particular, even the simplest version of an electron beam dump experiment like we describe may probe the range of kinetic-mixing parameters where a sub-GeV gauge bosons explains the $(g-2)_\mu$ anomaly \cite{Pospelov:2008zw}, while a lower-background experiment could probe kinetic mixing at the $10^{-4}$ level, well into the allowed region for unified theories, on a $\sim 1$-year timescale.  

\section{Simple Models of $\MeV-\GeV$-scale Dark Matter}
\label{Sec: Simple Models}
Some previous studies of invisibly-decaying $U(1)_D$ scenarios have taken \eqref{eq:lagrangian} or its scalar-DM counterpart to be the complete theory of dark matter \cite{deNiverville:2011it}. This approach is overly restrictive since, as noted above, more complex models of sub-GeV dark matter typically still have fixed-target physics governed by \eqref{eq:lagrangian} as an effective ``simplified model.''  In this section, we consider the physics of dark matter from a sub-GeV dark sector more generally.  For concreteness, we focus here on models where the cosmological abundance of dark-sector particles arises from thermal freeze-out, though models with non-thermal, matter-symmetric or -asymmetric abundances also exist (see \cite{Kaplan:1991ah,Kaplan:2009ag, Steffen:2006hw, Hamaguchi:2009hy, Petraki:2013wwa} and references therein).
 We will discuss cosmological and astrophysical constraints on these models in Sec.~\ref{ssec:supernova} and \ref{ssec:cosmoastro}.

The premise of a new, sub-GeV stable particle (SSP) immediately raises two questions: Why is it stable?  And how does it annihilate to a number density consistent with observations?  The cosmological stability of a sub-GeV particle $\chi$ may be ensured if it is the lightest particle charged under some global symmetry; one simple and motivated possibility is that it carries some charge under an unbroken sub-group of a spontaneously-broken gauge symmetry.  

If the SSP couples to ordinary matter with detectable strength, then it would have thermalized in the hot early Universe; this thermal abundance should be depleted, as the Universe cools, by efficient annihilation.  Annihilation mechanisms mediated by heavy particles of mass $M$ have cross-sections suppressed by $T^2/M^4$ at temperatures $T\ll M$, and therefore produce an excessively large SSP abundance.  This motivates theories involving at least one additional dark-sector particle $X$ --- perhaps a $U(1)$ factor of the gauge group motivated above, but possibly another scalar or fermion, which can decay into Standard  Model matter through new relevant or marginal (but naturally small) interactions.  

For simplicity we specialize to the case of a $U(1)_D$ gauge boson, in which case this minimal particle content is simply that of \eqref{eq:lagrangian}.  If $m_{A'}<m_\chi$ then $\chi\bar\chi \rightarrow A'A'$ annihilation proceeds with cross-section $\sim \pi \alpha_D^2/ m_{\chi}^2$ and the $\chi$ relic density is typically \emph{less} than the DM abundance (roughly $\Omega_{\chi}/\Omega_{DM} \sim 10^{-3} (\alpha / \alpha_D)^{2} (m_{\chi}/100 \ \MeV)^2$), and the $A'$ decays visibly.  If instead $m_{A'}>m_{\chi}$, the annihilation cross-section scales as $ \alpha_D \alpha\epsilon^2 m_{\chi}^2 /m_{A'}^4$ --- while this can give rise to a viable relic density for $\sim 1-10~\MeV$ $A'$ and $\chi$ or for large $\epsilon$ and $\alpha_D$ \cite{Pospelov:2007mp}, much of this parameter space for heavier $A'$ would over-produce dark matter.  

Thus, although corners of parameter space allow GeV-scale thermal dark matter, the more generic expectation is that the SSP $\chi$ is either parametrically over-produced (if it is lighter than the $A'$) or under-produced (if it is heavier).  The former case is inconsistent with observations; the second --- where $\chi$ is only a sub-dominant 
component of the dark matter --- is an interesting possibility that fits naturally into models where a \emph{heavy} stable particle that also carries dark-charge is \emph{the} dark matter (as in e.g.~\cite{Essig:2010ye}).  

It is still an interesting question whether a thermal GeV-scale particle can more naturally dominate $\Omega_{DM}$.  To realize this, it is useful to consider a slightly extended dark sector (but still far simpler than the Standard Model!) --- for example, one with at least two stable species and two dark-sector gauge interactions.  This larger model-space allows for annihilation that is slightly $A'$-mass-suppressed, but \emph{not} $\epsilon$-suppressed, whereas these two suppressions were artificially linked in the simple model based on \eqref{eq:lagrangian}.  
For illustration, consider a dark sector with gauge group $U(1)_h \times U(1)_l$ in a higgsed phase.
Both $U(1)$ gauge bosons can kinetically mix with the photon with coefficients $\epsilon_h$ and $\epsilon_l$, and we denote their masses and gauge couplings by $m_{A_h}, m_{A_l}$ and $\alpha_h, \alpha_l$, respectively. Consider also a Dirac fermion $\chi_h$ with unit charge only 
under $U(1)_h$ and a Dirac fermion $\chi_l$ with unit charge under both $U(1)$'s. For concreteness, let $m_{A_h} > m_{\chi_h} > m_{\chi_l} > m_{A_l}$.  
As mentioned above, the lighter state $\chi_l$ will naturally comprise a sub-dominant component of the dark matter via annihilations into $U(1)_l$ gauge bosons, 
which decay promptly to Standard Model leptons. The heavier state,
$\chi_h$, annihilates to $\chi_l$ with cross-section $\sigma_h \approx \pi \alpha_h^2 m_{\chi_h}^2 / m_{A_h}^4$.
The mild off-shell $A_h$ suppression of $\sim 10^{-3}$ makes it rather straightforward for $\chi_h$ to comprise the full relic density
for $\alpha_h$ in the $\sim 10^{-3} - 1$ range with $m_{\chi_h}$ in the $1-100\ \s \ \MeV$ range. 
Annihilations directly into Standard Model leptons are suppressed by $\epsilon_h^2\sim 10^{-6}$ relative to $\sigma_h$, and annihilations 
into $\chi_l$ do not induce charged particle production near recombination temperatures, so CMB constraints are satisfied. 
This model nicely accounts for all of the dark matter and is consistent with existing constraints (discussed further in Sec.~\ref{Sec: Constraints}). 
The fixed-target phenomenology may involve either (or both) $U(1)_{h,l}$, depending on $\epsilon_{h,l}$. If $\epsilon_l \ll \epsilon_h$, 
then the fixed-target physics involves only $U(1)_h$, and both $\chi_h$ and $\chi_l$ can be produced. If $\epsilon_l \gg \epsilon_h$, then the 
fixed-target physics only involves $U(1)_l$ and $\chi_l$, with $\chi_l$ as sub-dominant dark matter component. 

This discussion illustrates viable models of light dark matter for which the fixed-target phenomenology is well-described by the simplest 
$U(1)_D$ model with fermion or scalar $\chi$. This model can be compatible with CMB and galaxy constraints discussed in Section 
\ref{ssec:cosmoastro} over the full range of $m_{\chi_h}$, $m_{A'}$, $\epsilon$, and $\alpha_D=\alpha_h$ that we consider. Moreover, $\chi$ can be {\it the} dark matter, 
or a sub-dominant component, and it can be a fermion or scalar without loss of generality.
 Other scenarios can similarly be envisioned \cite{ArkaniHamed:2008qn, Pospelov:2007mp, Ho:2012br}, and in future work we will discuss 
aspects of these simple $\MeV - \GeV$ -scale dark matter models in more detail. 


\section{    Existing Constraints on $\MeV$-$10 \ \GeV$ Long-Lived Dark Sector Particles     }\label{Sec: Constraints}

This section summarizes present constraints on long-lived dark sector particles, which fall in two classes.  In the first class are those (from terrestrial experiments and supernova physics) that depend only on the $\chi$ interactions with matter and its approximate stability (typically $c\tau \gtrsim 1-1000\ \m$ depending on the experiment); these can be formulated as constraints on the parameters $\epsilon$, $m_{\chi}$, and $m_{A'}$ of \eqref{eq:lagrangian}.  To our knowledge, two important constraints of this type have not been considered in previous literature on these models: those from direct $\chi\bar\chi$ production at B-factories and from anomalous cooling of supernovae by $\chi\bar\chi$ production.  These new constraints are discussed in Sections \ref{ssec:collider} and \ref{ssec:supernova}.  Other laboratory constraints on this parameter space, including the $(g-2)_{e/\mu}$, $K^+\rightarrow \pi^+ + \rm{inv.}$, and $J/\Psi$ lines shown in the Figures, are summarized in Section \ref{ssec:otherlab}.

The second class of constraints, discussed in Section \ref{ssec:cosmoastro}, is more familiar to students of dark matter: bounds on $\chi$-electron scattering from direct detection experiments, and on $\chi$ self-interactions and $\chi\bar\chi$ annihilations into charged particles \cite{Lin:2011gj}.  These depend not only on the physics of \eqref{eq:lagrangian}, but also on the cosmological abundance of $\chi$ and, in the case of annihilation limits from the CMB and our Galaxy, on the dominant channels for $\chi\bar\chi$ annihilation. 
The $\sigma_{\chi e}$ and self-interaction limits can be re-interpreted as constraints on $\epsilon$ for given $m_{A'}$ , $m_{\chi}$, and $\alpha_D$ \emph{if} $\chi$ comprises the majority of dark matter, and are shown on the $\sigma_{\chi e}$ vs. $m_{\chi}$ plots under this assumption.  

The DM-annihilation limits are quite sensitive to the full model of light dark matter, as we illustrate by considering the $U(1)_l\times U(1)_h$ model elaborated in Section \ref{Sec: Simple Models}.  In particular, CMB observations rather severely exclude sub-GeV dark matter that has a matter-antimatter symmetric abundance and annihilates directly into SM-charged particles.  But these constraints are dramatically weakened if the dominant DM component $\chi_h$ annihilates instead to another dark-sector state, $\chi_l$, which in turn has visible annihilation products but low enough relic density that it is not significantly constrained.  Furthermore, asymmetric mechanisms for generating the dark matter density rather naturally yield viable dark matter scenarios.

\subsection{Electron Collider Constraints}\label{ssec:collider}

Two types of search at $e^+e^-$ colliders are sensitive to the $A'$-mediated production of $\chi \bar\chi$: searches for tagged mesons decaying invisibly (e.g. $\Upsilon(1S) \rightarrow invisible$ at BaBar \cite{Aubert:2009ae}) through the $A'$-mediated $b \bar b \rightarrow  \chi \bar\chi$ do not depend on the mass hierarchy between the $\chi$ and $A'$, but are relatively weak.  Greater sensitivity can be reached in searches for the continuum process $e^+e^- \rightarrow \gamma+ A'$ or $\gamma+\chi\bar\chi$ through an off-shell $A'$.  Although no search for this process has been published, one can extract a limit from a BaBar search for the decay $\Upsilon(3S) \rightarrow \gamma+ A^0$ with $A^0$ an invisible scalar \cite{Aubert:2008as}.  To our knowledge, this work is the first to extract a limit from \cite{Aubert:2008as} on the continuum process.  A crucial subtlety in this limit extraction stems the presence of a large continuum instrumental background ($e^+e^- \rightarrow \gamma \gamma$ where one photon goes un-detected) that is kinematically quite similar to sub-GeV $A'$ signals.  The similarity of this instrumental background to the signal of interest will likely prevent future searches from substantially improving the sub-GeV $A'$ bound, even with the much higher luminosities at a super-B-factory.
 
The search reported in \cite{Aubert:2008as} uses the photon energy distribution in single-photon events to set a limit on the $\Upsilon(3S) \rightarrow \gamma+A^0$, with $A^0$ decaying invisibly --- such a process would produce a signal of mono-energetic photons, with energy $E_\gamma = (m_{\Upsilon}^2 - m_{A'}^2)/(2 m_\Upsilon)$.  The resulting exclusion is mass-dependent, but in the range from $0.7-4\cdot 10^{-6}$ for $A^0$ masses below about 7 GeV.  As the dataset contains $122\cdot 10^6$ $\Upsilon(3S)$ events (approximately $25 \fb^{-1}$), this corresponds to sensitivity to roughly $100-500$ $\gamma+A^0$ events.  The overall efficiency for detection of these events is $10--11\%$ --- this accounts both for the acceptance of the angular selections ($\approx 37\%$ for $m_{A'} \ll 10 \GeV$) and for additional, non-geometric efficiencies, which we infer to be $\approx 27\%$ on average, and take to be roughly independent of geometry).  The initial $\Upsilon$ is not tagged, so the same analysis is sensitive to the continuum signal $e^+e^- \rightarrow \gamma+A'$, $A'\rightarrow invisible$.  The differential cross-section for $\gamma+A'$ production was calculated in \cite{Essig:2009nc} to be
\be
\frac{d\sigma}{d\cos\theta_*} = \frac{2 \pi  \alpha ^2 \epsilon ^2}{E_{cm}^2} \frac{1+\cos ^2 \theta_*}{\sin^2\theta_*}.\label{eq:dsigmaBabar}
\ee
Integrating over the angular acceptance $-0.31 <\cos\theta_* < 0.6$ of the BaBar search and using $E_{cm} = m_{\Upsilon(3S)}=10.3 \GeV$, and $\alpha(m_b) = 1/132$ and the $25 \fb^{-1}$ luminosity, we obtain an event yield of 
\be
N_{\gamma A'}  = 37 \cdot \left(\frac{\epsilon^2}{10^{-6}}\right) \label{ApyieldBabar} ~~,
\ee
within geometric acceptance.  To compare this to the $\Upsilon(3S)\rightarrow \gamma A^0$ branching fraction limits from \cite{Aubert:2008as}, one must multiply the limits by the number of $\Upsilon(3S)$'s in the data set and by the geometric acceptance for that signal.  The resulting bound is 
\be
\epsilon^2_{90\% U.L.} = 1.2 \cdot BR_{90\% U.L.}\qquad , \qquad m_{A'} \gtrsim 1 \GeV\label{BRtoEpsilon}
\ee
which varies from $0.85$ to $5\cdot  10^{-6}$ depending on the $A'$ mass.  This limit is dominated by statistical uncertainty in the background, which has a smooth, non-peaking distribution in the energy range of interest for $m_{A'} > 1 \GeV$.  We may therefore expect the limit on $\epsilon^2$ to scale with luminosity ${\cal L}$ as ${\cal L}^{-1/2}$ until a systematic limit is reached, so that a $50\ \ab^{-1}$ Belle-II dataset might improve these bounds by up to a factor of 45.  

A new complication arises for $m_{A'} \lesssim 1 \GeV$: the energy of the $e^+e^- \rightarrow A' \gamma$ photons in this scenario is separated from $m_{\Upsilon}/2$ by less than BaBar's photon energy resolution.  There is also a continuum background peaked at $E_{\gamma} = m_{\Upsilon}/2$: the process $e^+e^- \rightarrow \gamma \gamma$ where one $\gamma$ escapes detection.  The $A'$ signal is essentially indistinguishable from this background, at least for small enough $A'$ mass --- the rate and kinematics are essentially the same.  Indeed, the existing BaBar search \cite{Aubert:2008as} uses the $\gamma + invisible$ rate in an off-resonance dataset to normalize this background contribution ---  a procedure that would subtract away any sufficiently low-mass $A'$ signal along with the background (unlike the $\Upsilon(3S)$ decay signals for which the search was designed).  We can still infer a limit on the $\gamma+A'$ rate for low $A'$ mass, but must allow for the possibility that the vast majority of the events modeled as $\gamma\gamma$ background could in fact be a low-mass $A'$ signal.  The $\gamma\gamma$ background component was fit to $N_{\gamma \gamma} = 110\pm 46$ events in \cite{Aubert:2008as}; we take the high 1-sigma error bar of 156 events as a rough estimate of the $\gamma+A'$ limit for $m_{A'} \lesssim 1\GeV$.  Using the estimated $27\%$ \emph{non-geometric} efficiency, we infer a limit on $A'$ yield of 580 events, and hence $\epsilon^2 < 1.5\cdot 10^{-5}$ --- roughly a factor of four weaker than the naive limit obtained from \eqref{BRtoEpsilon} in this mass range.  

In contrast to the high-mass case, higher statistics alone would not meaningfully increase the sensitivity of this search to light $A'$ --- it is therefore meaningless to scale by ${\cal L}^{-1/2}$ in this case.  Instead, an improved search must measure the 2nd-photon veto inefficiency in another final state.  This approach is likely limited by systematic uncertainties --- it seems reasonable to expect future sensitivity in the neighborhood of $\epsilon^2 \sim 10^{-6}$, but not much better unless the background can be significantly reduced by tighter veto requirements.  The presence of a peaking background that is nearly indistinguishable from a sub-GeV invisible $A'$ signal underscores the need for complementary searches for light dark-sector particles, \emph{particularly} in this low-mass region.  

It is also worth noting that, even if the on-shell $A'$ is too light to decay to $\chi\bar\chi$, the virtual-$A'$ process $e^+e^-\rightarrow \gamma \chi\bar\chi$ could also be seen at $B$-factories, with a characteristic mass distribution $d\sigma/dm_{\chi\bar\chi}^2 \propto 1/m_{\chi\bar\chi}^2$ and an overall suppression by $\alpha_D/2\pi$ in cross-section, relative to \eqref{eq:dsigmaBabar}. For $m_{\chi}\ll \GeV$, this mass distribution is peaked at $m_{\chi\bar\chi}^2$ near zero, so that a $\chi\bar\chi$ yield of 580 events with $m_{\chi\bar\chi}^2 \lesssim \GeV^2$ is similarly excluded, corresponding roughly to a limit of 
$\epsilon^2 \frac{\alpha_D}{2\pi} \log(\GeV/m_{\chi}) \lesssim 1.5\cdot 10^{-5}$.  For $m_\chi\gtrsim \GeV$, the shape of the photon energy distribution in the non-resonant process differs considerably from that of the resonant process, so we do not infer a limit from \cite{Aubert:2008as}, but dedicated $B$-factory searches would certainly be sensitive to this process.

\subsection{Supernova Constraints}\label{ssec:supernova}
Another important constraint that rather robustly applies for $m_{\chi}\lesssim 100 \ \MeV$ can be derived from supernovae observations, 
and has not previously been considered in the literature in the context of sub-$\GeV$ dark matter charged under new gauge interactions (see \cite{Turner:1987by}
for similar constraints on axions).
Core collapse supernovae release energy in the form of $\MeV$ neutrinos that escape $\sim 10$ km out of the core. 
For $m_{\chi}\lesssim 100 \ \MeV$, $A'$-mediated reactions can produce $\chi$. If $\chi$ can free stream out of 
the core, then the production rate must be exceedingly small, which constrains $\epsilon$ (and $\alpha_D$)
as a function of $m_{A'}$. If the $\chi$ do not free stream, then as far as we know, no robust constraint can be derived. 

To estimate these supernovae limits, let's first compute the free streaming requirement. 
First consider the case where $m_{A'}>m_{\chi}$, in which case the $\chi$-nucleon scattering cross-section is
$\sigma\approx 4\pi\alpha\alpha_D\epsilon^2\frac{m_{\chi}^2}{m_{A'}^4}\sim 0.37 \times 10^{-37} \cm^2 (\frac{\alpha_D\epsilon^2}{10^{-10}})(\frac{m_{\chi}}{10 \ \MeV})^2(\frac{10 \ \MeV}{m_{A'}})^4$.
Core number densities of supernovae are $n_B\sim 1.7\ \cdot 10^{38} \cm^{-3}$ for densities of $\rho\sim 3\times 10^{14} \g/\cm^3$.
This gives a free streaming length of $l_{path}=1 / n_B \sigma \sim 10~{\rm km} (\frac{10^{-9}}{\alpha_D\epsilon^2})(\frac{20 \ \MeV}{E})^2(\frac{m_{A'}}{8 \ \GeV})^4$,
where $E$ is the $\chi$ energy. 
For the range of $m_{A'}$ and $\alpha_D\epsilon^2$ relevant for fixed-target searches, $\chi$ production does not free stream.
The free streaming regime is much more relevant for the higher mass range of $m_{A'}\sim 5 \ \GeV$ that B-factory searches can cover. 

To estimate the production rate, we'll follow the analysis of \cite{Dent:2012mx} using proton-neutron and proton-proton collisions 
and rescale the supernovae cooling bound from $\epsilon^2 \lesssim 10^{-20}$ to $\frac{\alpha_D}{4\pi}\epsilon^2(\frac{T}{m_{A'}})^4\lesssim 10^{-20}$ 
to account for off-shell production, as this is the regime where free streaming can occur.  
Plugging in $T\approx 30 \ \MeV$ we have $\alpha_D\epsilon^2\lesssim 6\times 10^{-10} (\frac{m_{A'}}{8 \ \GeV})^4$.
Thus, for the higher mass range of $m_{A'}\sim 5 \ \GeV$ that upcoming B-factory searches can cover, 
supernovae cooling would already constrain portions of the parameter space near $\alpha_D\epsilon^2\sim 10^{-9}$. 
This is of course for $m_{\chi}\lesssim 50-100 \ \MeV$, above which there is no cooling constraint. 
 At lower $m_{A'}\lesssim 100 \ \MeV$ where free streaming does not occur, it would still be interesting to investigate the extent to which 
supernovae dynamics can probe these scenarios, as the total production rate of $\chi$ is comparable to neutrino production.

\subsection{Other Laboratory Constraints} \label{ssec:otherlab}

Several other laboratory constraints on invisibly decaying $A^\prime$ have appeared in the literature.  
In Fig.~\ref{fig:SummaryReach} we show bounds from low energy probes of QED. For $m_{A^\prime}\lsim 30$ MeV, the dominant bound is
 from $A^\prime$ contributions to the electron-photon vertex. Since there is currently a $\sim 1.5 \sigma$ discrepancy between the SM prediction
 and the lower measured value of $(g-2)_e$  \cite{Giudice:2012ms}, the purple curve marks the parameter space for which $A^\prime$ corrections exceed
  the dominant {\it theoretical} uncertainty of $a_e \equiv (g-2)/2$ by $2 \sigma^{a_e}_{theory} = 1.6 \times 10^{-13}$. There are similar bounds from $(g-2)_\mu$, however, the discrepancy between theory and observation currently exceeds $3 \sigma$ so  the light green band in Fig.~\ref{fig:SummaryReach} shows where $A^\prime$ contributions 
 bring theory and experiment into $2 \sigma$ agreement \cite{Pospelov:2008zw}.  The dark green constraint marks where the disagreement exceeds $5 \sigma$. 
 
 For $m_{A^\prime}$ in the 30$-$300 MeV range with $2m_{\chi} < m_{A^\prime}$,
  the dominant constraint arises from rare kaon decays \cite{Artamonov:2008qb}. The brown curve in Fig.~\ref{fig:SummaryReach}
 uses measurements of the $K^+ \to \pi^+ \nu\bar \nu$ branching ratio to constrain the $K^+ \to \pi^+ + A^\prime$ width.
 For $2m_{\chi} < m_{K^+} - m_{\pi^+} < m_{A^\prime}$, there is also a constraint from the off shell $K^+ \to \pi^+ \chi\bar \chi$ decay, 
 but this is sensitive to $\alpha_D$ and further phase-space suppressed, so this constraint is weaker than the above-mentioned B-factory limits. In
 this regime, the parameter space that resolves the $(g-2)_\mu$ anomaly is largely unconstrained.  
 For $2m_\chi > m_{A^\prime}$, the $A^\prime$ decays visibly to 
 $e^+ e^-$, so the main constraint arises, instead, from $K^+ \to \pi^+ \ell^+ \ell^-$ decays. 
 
 Limits on $J/\Psi$ decays to $\chi\bar\chi$ \cite{Ablikim:2007ek}, on non-resonant effects of a kinetically mixed $A'$ \cite{Hook:2010tw}, which do not depend on its decay modes, and on resonant $A'$ production with decay into $e^+e^-$ or $\mu^+\mu^-$ final states \cite{Merkel:2011ze, Abrahamyan:2011gv, Aubert:2009cp, Echenard:2012hq, Babusci:2012cr, Adlarson:2013eza} have also appeared in earlier literature.   The first of these is relevant to the range of $\epsilon$ we consider only if the $A'$ mass is near that of the $J/\Psi$, in which case their mixing is resonantly enhanced.  The model-independent constraints of \cite{Hook:2010tw} constrain $\epsilon^2 \lesssim 10^{-3}$ for $A'$ lighter than the $Z$ boson; though slightly above the range considered in our plot, these become a leading constraint on the models considered here for $A'$ masses above $7-8$ GeV, where B-factory searches are ineffective.  

Finally, the $A'$ visible decay searches noted above are sensitive to $\epsilon^2$ as low as $10^{-6}$ depending on $A'$ mass (with significant improvements anticipated from future searches).  When other $A'$ decay modes are accessible, they reduce the visible decay signals constrained by these searches (so that they scale as $\epsilon^4\alpha/\alpha_D$ rather than $\epsilon^2$) while increase the $A'$ width (which exceeds the percent-level resolution-limited widths assumed by these experiments whenever $\alpha_D \gtrsim \alpha$).  Accounting for both effects, none of these searches imply bounds stronger than $\epsilon^2 \sim 10^{-3}$ for the models considered here.  For similar reasons, the beam-dump limits on weakly coupled $A'$ (see \cite{Bjorken:2009mm}, \cite{Andreas:2012mt}, and references therein), which rely on an $\epsilon^2$-suppressed $A'$ width and consequently long $A'$ lifetime, do not apply.

\subsection{Cosmological and Astrophysical Constraints}\label{ssec:cosmoastro}
Several further constraints rely on the relic abundance of $\chi$.  
Constraints from galactic halo structure on dark matter self-interaction constrain models with $\sim \MeV$-scale $\chi$ and $A'$ masses 
\cite{Lin:2011gj,Markevitch:2003at,*MiraldaEscude:2000qt}.
Following the discussion of \cite{Lin:2011gj}, dark matter self-interactions are constrained at the level of $\sigma_{DM} / m_{\chi} \lesssim (0.2-2) b/\GeV$. 
This can readily constrain parameters in the simple model of Section \ref{Sec: Simple Models} for low $m_{\chi}\sim 10 \ \MeV$ masses. 
For $m_{A'}\lesssim m_{\chi}$, $\sigma_{DM}\approx \pi \alpha_D^2 / m_{\chi}^2$, while for $m_{A'}\gsim m_{\chi}$ there is an 
extra suppression of $(m_{\chi}/m_{A'})^4$. 
We can interpret the bounds on the self-interaction cross-section as a limit on $\alpha_D$ for a given $m_{\chi}$ and $m_{A'}$.
In particular, we find $\alpha_D^2 \lesssim 2 \left(m_{\chi}/10 \ \MeV\right)^2 \left(m_{A'}/m_{\chi}\right)^4$ for $m_{A'}\gsim m_{\chi}$.
This constraint is not explicitly shown on our plots as it is satisfied for essentially all of the parameter range shown. 

Additional constraints on light dark matter derived from CMB and galactic observations are well known. 
The robustness of these constraints is sometimes overstated. Here we review 
them and show that the simple dark matter models of Section \ref{Sec: Simple Models} with $\GeV$-scale gauge 
forces are compatible with these considerations over the entire natural range of parameters relevant for the fixed-target phenomenology. 

For the sub-$\GeV$ -scale dark matter we're considering, late time annihilations into charged leptons 
are constrained by measurements of the CMB \cite{Galli:2011rz,*Hutsi:2011vx,*Finkbeiner:2011dx}. 
Taken from \cite{Galli:2011rz,*Hutsi:2011vx,*Finkbeiner:2011dx}, 
$(\frac{\Omega_{\chi}}{\Omega_{DM}})^2 \frac{\langle \sigma v\rangle}{m_{\chi}} \lesssim 5 \times 10^{-28} \cm^2\s^{-1} \GeV^{-1}$ 
is required by Planck measurements of the CMB. 
For $\Omega_{\chi}=\Omega_{DM}$, this translates into $\sigma_{leptons} < (\frac{m_{\chi}}{\MeV}) 10^{-5} \sigma_{thermal}$.
For sub-dominant components, we can use $\Omega\sim 1 / \langle \sigma v\rangle$ to write the constraint as
$(\frac{\Omega_{\chi}}{\Omega_{DM}}) \lesssim (\frac{m_{\chi}}{100 \ \MeV}) 10^{-3}$.

Turning back to the examples discussed in Section \ref{Sec: Simple Models}, $\chi_l$ components will parametrically 
have a sub-dominant relic density of order $(\frac{\Omega_{\chi}}{\Omega_{DM}}) \sim 10^{-3}(\frac{\alpha}{\alpha_D})^2(\frac{m_{\chi}}{100 \ \MeV})^2$,
so CMB constraints are naturally satisfied for typical model parameters. 
For the $U(1)_h\times U(1)_l$ scenario, the thermal relic density is set by the cross-section for $\chi_h$ annihilations
into $\chi_l$. $\chi_h$ annihilates to leptons with cross section $\epsilon_h^2\sim 10^{-6}$ smaller, 
so again CMB constraints are satisfied. The lighter $\chi_l$ component is sub-dominant. 

Annihilations into charged leptons at low energy are also constrained by INTEGRAL/SPI measurements 
of the flux of $511~\keV$ energy photons from the galactic center \cite{Prantzos:2010wi,*Bouchet:2010dj}. 
The requirement that $\chi$ annihilations not inject charged leptons at a rate larger than the measured flux 
amounts to \cite{deNiverville:2011it} $\langle\sigma v\rangle \lesssim 10^{-4}~ {\rm pb}\cdot  c (\frac{m_{\chi}}{\MeV})^2(\frac{\Omega_{DM}}{\Omega_{\chi}})^2$.
This constraint is somewhat weaker than the CMB constraints, but scales differently with mass $m_{\chi}$.
For the example models, this requirement is naturally satisfied. 
Likewise, constraints on annihilations into charged leptons from measurements of the photon spectrum in the galactic center 
 are weaker than CMB constraints. 

Other constraints on light dark matter can be derived from measurements of the number of effective neutrino species \cite{Boehm:2013jpa}. 
For $m_{\chi} > 3 \ \MeV$ in the range considered in this paper, this condition is met.

\section{    Signal Production and Detection     }\label{Sec: ProductionDetection}

A multi-GeV electron beam impinging on material, as in a beam dump, loses energy primarily through bremsstrahlung in coherent electron-nucleus scattering.  
A fraction of the electrons (of order $\alpha m_e^2/m_\mu^2$) exchange sufficient momentum with the nucleus to pair-produce muons. Similarly, any non-SM matter that interacts with electrons or photons will also be produced in radiative processes, with a rate suppressed by its coupling and mass. If, as in the case of the $\chi$ particles of  \eqref{eq:lagrangian}, the particles thus produced are long-lived and penetrating, they can be observed through their scattering in a detector downstream of the dump. 

For the model of \eqref{eq:lagrangian}, the leading production and scattering processes are illustrated by Figure \ref{fig:prod}, where the gauge boson $A'$ may be on- or off-shell, depending on the particle masses.
A complete Monte Carlo model of $\chi$ production and scattering, described in the Appendix, has been used to generate all sensitivity plots.  
This section summarizes the essential physics of the production and scattering processes, making several approximations to more simply illustrate the scaling and typical kinematics of the reactions.  The key results of each subsection, which allow the reader to scale our results to other beam energies and detector geometries, are eqns. \eqref{Apyield}, \eqref{offshellYield},\eqref{acceptance}, and \eqref{scatteringProb}. These simplified formulas reproduce the results of the full Monte Carlo to within a factor of 2.  
The following sub-sections describe the $\chi$ yield per incident electron, the characteristic $\chi$ kinematics and resulting geometric acceptance for a distant detector, and the elastic $\chi$-nucleon and $\chi$-electron scattering rates in a detector.  

\subsection{$\chi \bar \chi $ Production in an Electron Beam-Dump}\label{ssec:production}

The cross-section and characteristic kinematics for  $\chi \bar\chi$ production in scattering of an electron beam of energy $E_0$ on a nuclear target of atomic number $Z$ can be simply approximated in the Weizsacker-Williams approximation.  We consider first the case of production of an on-shell $A'$ (``$A^\prime$-sstrahlung" ) that subsequently decays to a $\chi\bar\chi$ pair, then extend this result to the case of $\chi\bar\chi$ pair production mediated by an off-shell $A'$. 

The differential  $A^\prime$ electro-production cross-section was computed in \cite{Bjorken:2009mm} in the Weizsacker-Williams approximation as 
\begin{align}
&\frac{d\sigma_{eN\rightarrow eNA'}}{dx d\cos\theta_A^\prime} \approx (8  \alpha^3 \epsilon^2) \, \Phi(q_{min},q_{max})\frac{E_0^2 x}{ U^2}  \nonumber \\ 
& \times \left[   \left(1-x+\frac{x^2}{2}\right)    - \frac{x^2 (1-x) m_{A^\prime}^2  (E_0^2 \theta^2_{A^\prime}+m_e^2)  }{U^2} \right] , \\  \nonumber\\
& U(x, \theta_{A^\prime}) =  E_0^2 x \theta^2_{A^\prime} + m_{A^\prime}^2 \frac{1-x}{x} + m_e^2 x,\\
& q_{min} = \frac{2 U}{E_0(1-x)} \qquad q_{max} = m_{A'},
\end{align}
where $Z$ is the target nucleus' atomic number, $\theta_{A^\prime}$ is the angle between $A^\prime$ and the beam axis
in the lab frame, $x=E_{A'}/E_0$ the fraction of the beam energy carried by the $A'$, and $\alpha\Phi(q_{min},q_{max})/\pi$ is the Weizsacker-Williams effective photon flux \cite{Kim:1973he} for photons with virtuality $q_{min}^2 < -t < q_{max}^2$, discussed further in Appendix \ref{appss:production} (the function we call $\Phi$ is usually denoted as $\chi$, but we have changed notation to avoid confusion with the particle $\chi$). For $q_{min}$ much larger than the inverse nuclear size $\approx 0.4 \GeV/A^{1/3}$, $\Psi(q_{min},q_{max})$ is proportional to $Z^2$ times a logarithmic factor.  

\begin{figure}[t]
 \vspace{0.1cm}
  \includegraphics[width=8.6cm]{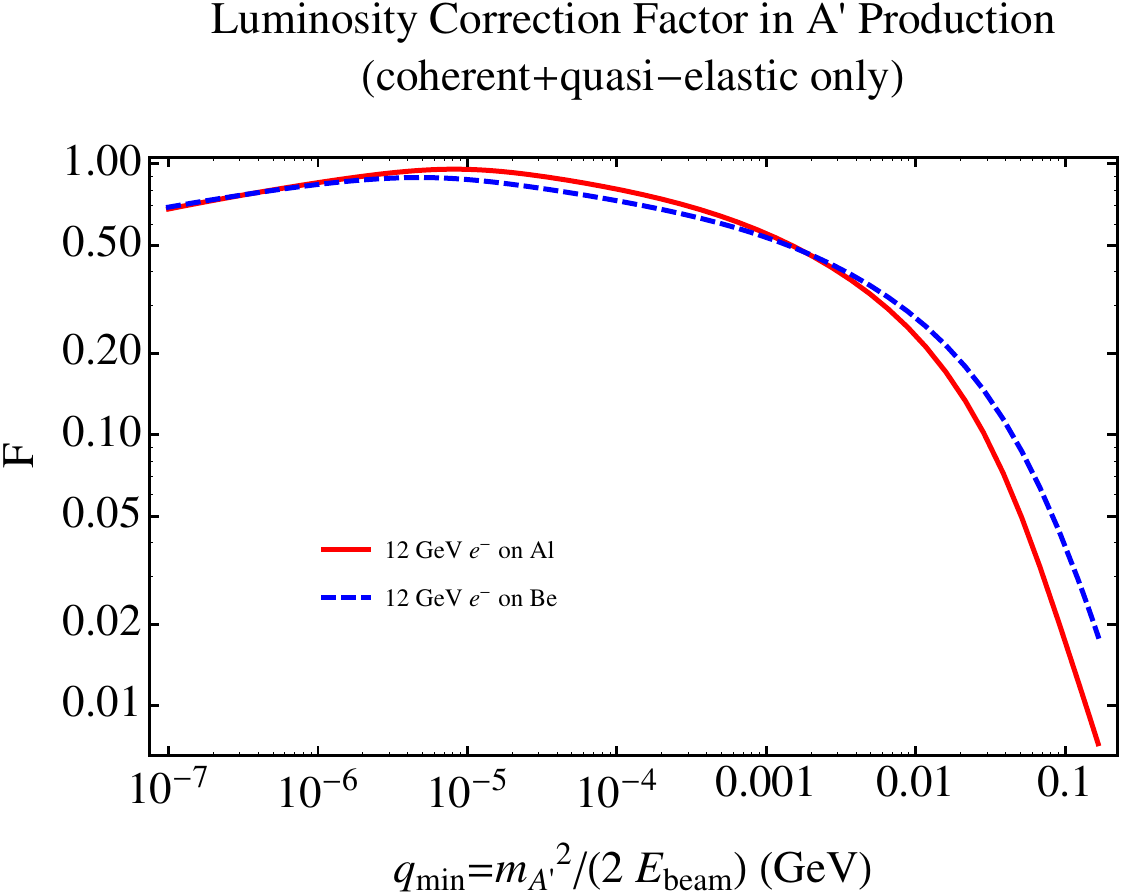}
  \caption{
The ``luminosity correction'' factor $F(q_{min}=m_{A'}^2/(2 E_{beam}),q_{max}=m_{A'})$ defined in \eqref{fdef}, for use in estimating $A'$ yield. $F$ is proportional to the Weizsacker-Williams effective photon flux $\Phi(q_{min},q_{max})$, multiplied by a material-dependent luminosity factor.  
The red solid and blue dashed curves correspond to 12 GeV electron beams impinging on a thick Aluminum or Beryllium target (i.e. $q_{max}= \sqrt{2\cdot 12\GeV\cdot q_{min}}$).
For the $A'$ mass ranges of interest, $F$ can depend sensitively on $q_{min}$ but far less on $q_{max}$, so that these curves remain approximately valid for other beam energies.  For example, scaling beam energy and $m_{A'}^2$ simultaneously by up to a factor of 10 (not shown), so that $q_{min}$ is unchanged but $q_{max}$ scales by $\sqrt{10}$, only affects $F$ at the $\sim 20\%$ level or smaller.  The expression for $F$ shown here includes \emph{only} coherent elastic nuclear scattering and quasi-elastic scattering off nucleons; neglecting inelastic contributions may underestimate the $F$'s for 1--2 GeV $A'$ masses at the highest $q_{min}$ shown by a factor of $\sim 5$.}\label{fig:coherence}
\end{figure}

As was noted in \cite{Bjorken:2009mm}, for any given $x$ the angular integral is dominated by angles $\theta_{A'}$ such that $E_0 x \theta_{A'}^2 \lesssim  m_{A^\prime}^2 \frac{1-x}{x} + m_e^2 x$. Neglecting $O(m_e^2)$ terms and the angular dependence of $\Phi$, we obtain a simple approximate differential cross-section 
\be
\frac{d\sigma}{dx} = (4\alpha^3 \epsilon^2 )\bar\Phi(m_{A'},E_0) \frac{x^2 + 3 x(1-x)}{3(1-x)m_{A'}^2} ~~ ,\label{dsigmadx}
\ee
where $\bar\Phi(m_{A'},E_0) \equiv \Phi(q_{min} = m_{A'}^2/(2 E_0), q_{max} = m_{A'})$.  
This expression is in turn dominated at $1-x\ll 1$ (the apparent log divergence as $x\rightarrow 1$ in \eqref{dsigmadx} is regulated by $O(m_e^2)$ terms and corrections to the Weizsacker-Williams approximation when $1-x \lesssim \delta$ with
\be
\delta \equiv \max(m_{A'}/E_0, m_e^2/m_{A'}^2, m_e/E_0).
\ee
The total $A'$ production cross-section scales like 
\be
\sigma_{A'} \approx \tfrac{4}{3} \frac{\alpha^3 \epsilon^2}{m_{A'}^2} \bar\Phi(m_{A'},E_0)   \left[   \log(1/\delta)+O(1) \right]~~ . \label{crossSectionWW}
\ee
The $A'$ yield for a mono-energetic beam on a target of $t$ radiation lengths is given by 
\be
N_{A'} = \sigma_{A'} \cdot \pf{t X_0 N_0}{A},
\ee
where $X_0$ is the radiation length of the target in $\g/\cm^2$, $A$ the atomic mass in $\g/\rm{mole}$, and $N_0$ Avogadro's number (the latter factor is the ``luminosity'' of nuclei encountered per incident electron).  As discussed in Appendix \ref{appss:production}, for thickness $\gg 1$ the contributions from a degraded beam in a thick target may be conservatively modeled by using this mono-energetic formula with $t\rightarrow 1$.  
Since the radiation length is itself determined by electromagnetic processes, it is useful to introduce the combinations
\be
F(q_{min},q_{max}) \equiv \frac{4}{3} \alpha^3 \Phi(q_{min},q_{max}) \frac{X_0 N_0}{A m_e^2}
\label{fdef}
\ee
and $\bar F(m_{A'},E_0)$ defined similarly in terms of $\bar\Phi$.  All target-dependence in the yield is absorbed into this ``luminosity correction'' factor $F$, which is 
$O(1)$ over most of the parameter range of interest (see Figure \ref{fig:coherence}) regardless of the target nucleus, with suppression only when $q_{min}$ becomes comparable to the inverse nuclear size.  In terms of $F$, the $A'$ yield from a thick target is reasonably approximated by
\begin{align}
N_{A'} \approx N_e &\bar F(m_{A'},E_{beam})\cdot\pf{m_{e}^2}{m_{A'}^2} \cdot \epsilon^2 \label{Apyield}\\
&\times \log\left[\min(E_{beam}/m_{A'},m_{A'}^2/m_e^2)\right].\nonumber
\end{align}
The factor $\bar F$ falls at high $m_{A'}$ because of two comparably important effects: the loss of nuclear coherence for $q$ comparable to the inverse nuclear size and the shrinking range of integration in \eqref{ChiExp}.
The expression for $F$ shown in \ref{fig:coherence} includes \emph{only} the contributions from coherent elastic nuclear scattering and quasi-elastic scattering off nucleons described in Appendix \ref{appss:production}; rough estimates suggest that, at the highest $q_{min}$ shown, inelastic scattering off nucleons increases $F$ by a factor of up to $\sim 5$ relative to the curves shown.

Because the distribution of $x$ is log-divergent in the kinematic region the Weizsacker-Williams approximation breaks down, we can predict only qualitative features of the $A'$ energy and angular distributions --- but because we are interested here in the kinematics of a single $A'$ decay \emph{product}, the two essential features are:
\begin{itemize}
\item The $A'$ \textbf{energy} is peaked at $x \approx 1$, with ${\rm median}(1-x)\sim O(\sqrt{\delta})$.   From full simulation, we find $0.02 < (1-x)_{median} < 0.2$ for MeV--GeV $A'$ produced from a 12 GeV beam).  
\item The $A'$ \textbf{angle} relative to the beam-line is also peaked forward (roughly as $m_{A'}/E \times \delta^{1/4}$) in a narrower region than the typical  opening angle for the $A'$, i.e. $m_{A'}/(E_0 x)$.   In the region where the detector acceptance differs significantly from unity, $m_{A'}\gtrsim  200 \, \MeV$, the median $A'$ production angle is $0.1-0.2$ times the decay angle. 
\end{itemize}
In light of these features, a simplified picture of $\chi$ production in which we treat the $A'$ to be produced strictly forward with energy $E_{A'} \approx E_{0}$ is approximately valid. 
In a very real sense, a secondary {\it beam} of $\chi$ particles is produced by the primary electron beam with very sharply peaked forward kinematics. 

By contrast, the energy and angular distribution for $A'$ production (with $m_{A'}\lesssim m_p$) off a proton beam looks much more like familiar photon bremsstrahlung with lower median energy ($x_{med}\sim 0.1$) and subsequently larger $A'$ decay opening angles. 
In the case of $\chi$ produced in decays of secondary mesons from proton-beam interactions, the median energy fraction carried by $\chi$ is still typically in the $x\sim 0.05-0.1$ range, leading to a rather un-collimated secondary beam of $\chi$s. 

The qualitative features of on-shell $A'$ production off an electron beam apply equally to the case of $\chi\bar\chi$ production mediated by an off-shell $A'$.  
Indeed, the cross-section differential in $\schi \equiv (p_\chi+p_{\bar\chi})^2$ can be written simply as
\be
\frac{d\sigma}{d\schi d\dots} = \frac{d\sigma_{A'}(\sqrt{\schi})}{d\dots} \times \frac{1}{\pi}\frac{\sqrt{\schi} \Gamma_{A'\rightarrow \chi\bar\chi}}{|s-m^2 + \sqrt{\schi} \Gamma_{A'}(\sqrt{\schi})|^2} \nonumber
\ee
where $ \frac{d\sigma_{A'}(\sqrt{\schi})}{d\dots}$ is a (possibly differential) cross-section for on-shell $A'$ production with $m_{A'} \rightarrow \sqrt{\schi}$ and $\Gamma_{A'\rightarrow \chi\bar\chi}(\sqrt s )$ is the partial width of a would-be $A'$ of mass $\sqrt{s}$, i.e. for fermionic $\chi$ 
\be
\sqrt{s} \Gamma_{A'\rightarrow \chi\bar\chi}(\sqrt s) = \frac{\alpha_D s}{3} \sqrt{1-4 y} (1+2y)\quad y=m_\chi^2/s. \nonumber
\ee

The above formula manifestly has the right propagator form, and reproduces \eqref{dsigmadx} for the resonant contribution.
Far above the $A'$ resonance, this gives
\be
\frac{d\sigma}{d\schi} = \frac{(4\alpha^3 \epsilon^2 )\Phi}{\schi} \log(1/\delta) \times \frac{\alpha_D}{3 \pi} \frac{\sqrt{1-4 y} (1+2 y)}{\schi}.\label{ddschi}
\ee 
The $1/s^2$ cross-section implies that the production is dominated near threshold, at $\sqrt{\schi}\sim (2-4) m_{\chi}$.   The peaking of the angle--energy distribution at forward angles and high $\chi\bar\chi$ pair energy that were noted above continue to hold, with the role of $m_{A'}$ now played by $(few)\times m_{\chi}$.  A reasonable approximation to this scaling in the case of fermionic $\chi$ is 
\be
N_{\chi\bar\chi} \approx \pf{\alpha_D}{\pi} N_{A'}\bigg|_{m_{A'} = \sqrt{10.} m_{\chi}},\label{offshellYield}
\ee
where the second factor denotes the result of \eqref{Apyield} at the fictitious $A'$ mass that dominates the $\schi$ integral.  For bosonic $\chi$ produced through an off-shell $A'$, the differential cross-section analogous to \eqref{ddschi} is $p$-wave suppressed near threshold, resulting in a further suppression of yield by roughly an order of magnitude.  

\subsection{Geometric Acceptance}
In the $\chi\bar\chi$ center-of-mass frame ($A'$ rest frame for the on-shell case), 
the $\chi$ will be produced with an angular distribution $dN/d\cos\theta_* \propto 1\pm \cos\theta_*^2$ with the positive sign for relativistic fermonic $\chi$ and negative for relativistic bosonic $\chi$.  Boosted into the lab frame, the typical opening angle is $\theta_\chi \approx m_{A'}\beta_{\chi}/E_0$ where $\beta_\chi = \sqrt{1-4m_\chi^2/m_{A'}^2}$ is order-1 except for near-threshold decays.   
As above, for  $\chi\bar\chi$ production through a virtual $A'$, the typical invariant mass of $(2-4) \cdot m_{\chi}$ can be substituted for $m_{A'}$.    
When $m_{A'}/E$ is smaller than the angular size of the detector, the angular acceptance is $O(1)$ for both bosonic and fermonic $\chi$.  More generally, the acceptance for fermionic $\chi$ is well approximated by
\begin{align}
&(1+m_{A'}^2/\theta_D^2E_0^2)^{-1}  &\hspace{-28pt}\mbox{on-shell $A'$} & \\
&(1+m_{\chi}^2/\theta_D^2E_0^2)^{-1}  &\hspace{-28pt} \mbox{off-shell $A'$}&,\label{acceptance}
\end{align}
where $\theta_D$ is the angular size of the detector, i.e. for a detector of diameter $L_d$ at distance $d$ from the dump, $L_d/(2 d)$. 
In the case of bosonic $\chi$ (not considered in detail in this work) with $m_{\chi} \ll m_{A'}$ and $m_{A'}/E_0 \gtrsim \theta_D$, further suppression arises from the angular distribution in $A'$ decay, which has a node in the forward direction.  For the geometry we consider, this effect lowers the acceptance in $\chi\bar\chi$ production by a factor of up to $\sim 3$ at $A'$ masses above $\sim 500$ MeV.

\subsection{Scattering Signals and Total Yields}

The primary signal considered in this paper is quasi-elastic $\chi$-nucleon scattering with momentum 
transfer $Q^2 > (140 \,\MeV)^2$.  This range corresponds to nucleon recoil energies above 10 MeV, above ``fast'' neutron and radiological backgrounds.
Although there are also contributions from coherent $\chi$-nucleus interactions, 
inelastic scattering, and electron recoils, we conservatively consider only the quasi-elastic nucleon signal for simplicity. This process
is already used to study neutral-current interactions at MiniBooNE \cite{Perevalov:2009zz} and is simpler to model numerically, but by ignoring other contributions, our results 
generically underestimate the projected sensitivity to new physics.

The typical $\chi$ produced by the electron beam has energy $E\sim E_{0}/2 \gg  m_{N},m_{\chi},\sqrt{Q^2}$.   In this limit, the scattering rate is given by
\be
\frac{d\sigma}{dQ^2} = (4 \pi \epsilon^2\alpha\alpha') \frac{F_{1,N}^2 - \frac{Q^2}{4m_N^2}F_{2,N}^2(Q^2)}{[m_{A}^2 + Q^2]^2},\label{eq:nucrecoilSimple}
\ee
where the nuclear monopole and dipole form factors are respectively 
\be
F_{1,N}(Q^2) &=& \frac{q_N}{(1+ Q^2/m_N^2)^2} ~~, \\ 
F_{2,N}(Q^2)  &=&  \frac{\kappa_N}{(1+ Q^2/m_N^2)^2}~~,
\ee
with $q_p=1, q_n = 0$ and $k_p = 1.79$ and $\kappa_n = -1.9$ \cite{AguilarArevalo:2010cx}.   
The inclusive nucleon-averaged scattering rate with $Q^2_{min}<Q^2 < Q^2_{max}$ for very light or heavy $A'$ is of order
\be
\sigma_{\chi N} \sim (4\pi \epsilon^2 \alpha \alpha') \frac{Z}{A}  \frac{1}{\mu^2} \qquad\qquad\qquad \\
1/\mu^2 \equiv \begin{cases} 
(Q_{max}^2 - Q_{min}^2)/m_{A'}^4 & m_{A'}^2 \gg Q^2_{max}\\
1/m_{A'}^2 & Q^2_{min}\lesssim m_{A'}^2\lesssim Q^2_{max} \\
1/Q_{min}^2 & m_{A'} \ll Q^2_{min}. \label{mu2defn}
\end{cases}
\ee
We neglect detection efficiencies, which for carbon-based detectors are typically near 100\% over the $Q \sim $100 MeV $-$ GeV range of interest \cite{AguilarArevalo:2010cx}. The probability that a single incident $\chi$ scatters with observable $Q^2$ in a detector of thickness $L_d$ and density $\rho$ is therefore
\be
P_{scat} &= &\frac{\rho}{m_N} L_d \sigma_{\chi N}\label{scatteringProb}
\\  \approx \alpha_D 10^{-7} &\times& 
\pf{\rho}{1 \g/\cm^3}\pf{D}{1\m}\pf{\epsilon}{10^{-3}}^2\pf{0.1 \GeV}{\mu}^2.\nonumber
\ee
Combining this result with the $\chi$ yield from \eqref{Apyield} or \eqref{offshellYield} and the angular efficiency penalty (\eqref{acceptance} with $\theta=D/(2\ell)$) gives a simple estimate for the total yield from any variation of the experimental scenario considered here; for the parameter ranges we have considered, the estimated result agrees with a detailed simulation to within a factor of 2.
To illustrate this estimation, we consider benchmark points $\epsilon^2=1.5\cdot 10^{-7}$ and $m_{A'} = 100 \ \MeV$ ($\epsilon^2=10^{-5}$ and $m_{A'} = 500 \ \MeV$) with 10 MeV $\chi$ and $\alpha_D=1$ (both chosen to lie near the 1000 event line in the upper-left panel of Figure \ref{fig:SummaryReach}).  The expected $A'$ yield from \eqref{Apyield} per $10^{22}$ electrons on target is $0.8\cdot 10^{12}$ ($0.9\cdot 10^{11}$ for the high-mass point).  For a 1 m square detector situated 20 meters from the dump, $\theta_d \approx 0.025$ so that the lower mass point has roughly 90\% acceptance and the higher mass-point roughly $25\%$.  Finally, $\chi$ incident on the detector have a $\chi$-nucleon scattering probabilities in a 1 m long mineral oil detector ($\rho \approx 0.8 \ \g/\cm^3$) from \eqref{scatteringProb} are $0.8\times 10^{-7}$ ($3\times 10^{-8}$).  Multiplying these factors we estimate $1250$ events for the 100 MeV point and 600 for the 500 MeV point --- both quite close to the 1000 events obtained by a full Monte Carlo.

We will not discuss $\chi$-electron scattering in the main results of this paper. However, we note for completeness that the $\chi$-electron scattering cross-section can be considerably larger for light $A'$. Up to corrections of ${\cal O}(m_e^2)$, the recoil profile for $\chi e \to \chi e$ scattering in the lab frame is
\be
 \frac{d\sigma_{\chi e}}{dE_{f}} = 
  4 \pi \epsilon^{2} \alpha \alpha^\prime m_e 
 \frac{  
 4 m_e m_\chi^2 E_f + [m_\chi^2 + m_e (E -E_f) ]^2
 }{    (m_A^2  +    2m_e E_{f} )^2 ( m_{\chi}^{2} + 2  m_{e} E)^2  } ,~
\label{eq:erecoil}
\ee
where $E$ is the incoming $\chi$ energy, and $E_f$ is the electron recoil energy; this formula applies when $E\gg m_e,m_\chi$. For a full treatment of electron scattering
see Appendix \ref{appss:erecoil}.

\section{Experimental Approach}\label{Sec: ExperimentalApproach}

The experimental approach combines two techniques that each have an illustrious history: 
using an electron beam-dump to obtain a secondary beam of light states and detection of quasi-elastic (elastic) neutral-current scattering off nucleons (electrons).  Scintillator detectors that can detect charged particles downstream of electron beam dumps have been successfully used in searches for axions and millicharged particles \cite{slac-mQ,*SLAC-thesis}. 
Electron beams were particularly useful in these cases due to their relatively low backgrounds compared to proton beams.  
Likewise, detection of neutral-current scattering behind specially designed proton beam dumps 
has formed the basis for modern precision neutrino physics.  

The combination of these two techniques --- detection of neutral-current signals in a small ($1~\m^3$-scale fiducial volume) detector downstream of an electron beam dump --- offers 
a powerful low-background method to search for $\MeV - \GeV$-scale dark matter, drawing on a wealth of past experience. The background rates and signal efficiencies of near-surface neutrino detectors like MiniBooNE \cite{MiniBooNE_A,*MiniBooNE_B,Perevalov:2009zz}, together with measurements at other near-surface facilities like CDMS-SUF \cite{TAP_THESIS,Chen} and the SLAC millicharge (mQ) search \cite{slac-mQ,*SLAC-thesis}, serve as a guide for understanding background sources in these experimental setups.

Relativistic dark matter produced in the beam dump can scatter off detector nuclei, nucleons, or electrons, with each reaction inducing a distinct signature. 
The largest process (for sufficiently light mediators) is coherent elastic nuclear scattering --- this can have energy deposition above radiological backgrounds, but suffers from 
potentially large slow neutron backgrounds from the beam dump and ``skyshine'' neutron backgrounds (neutrons that re-scatter from the atmosphere).
Quasi-elastic scattering off nucleons with $Q^2\gsim (140 \ \MeV)^2$ induces recoil energies above the slow neutron background; so long as higher-energy neutrons are ranged out, 
muons and neutrinos are the main beam-related background to this signal.
Above this energy range, a variety of inelastic reactions produce $\pi^+$ and $\pi_0$ in dark matter - nuclear collisions; the rates for these processes are somewhat lower, but if they can be efficiently discriminated from backgrounds then searches for these reactions could be quite powerful.  
Our focus in this work will be on quasi-elastic nucleon scattering.  We leave a broader study of inelastic, nuclear-elastic, and electron scattering backgrounds for future work, but stress that searching in multiple scattering channels may well give the best sensitivity over a wide range of mediator masses.

The kinematic features of electron beam fixed-target scattering are particularly well-suited 
for producing a collimated secondary beam of light dark matter, as described in Sec.~\ref{Sec: ProductionDetection}. 
The tight collimation allows the use of a small detector of $\mathcal{O}(1 \text{m})$ in size. 
Absolute neutrino rates are low enough that a detector can be situated close to the dump, thereby maintaining good angular 
acceptance even with a small detector. Muons and high energy neutrons can be ranged out (or deflected for muons) on a similar short distance 
scale for the energies under discussion. The resulting beam-related backgrounds, described in detail in Sec.~\ref{SubSec: BeamRelatedBackgrounds}, 
are negligible for this type of setup --- this is reinforced by past experience with mQ \cite{slac-mQ,*SLAC-thesis}.

The dominant potential backgrounds for high-$Q^2$ nucleon recoil signals are neutrons produced by interactions of cosmic-ray muons (cosmogenic neutrons). 
Other sources of background like radioactive decays, slow neutrons from the beam dump, and skyshine neutrons are significantly less energetic than dark-matter-induced recoils. On a one-year time scale, the flux of cosmogenic neutrons (above a 10 MeV energy threshold) on a $1\ \m^2$ detector can be as high as $\sim 5\times 10^5$. 

The primary way of reducing this background in existing near-surface neutrino detectors is by the use of
timing cuts that exploit the bunched structure of the beam \cite{MiniBooNE_A,*MiniBooNE_B, Perevalov:2009zz} to achieve  $10^{21-22}$ protons on target in a lifetime of only $10^{3-4} \s$.
The residual beam-unrelated backgrounds can be measured during beam-off periods, providing excellent control on systematic uncertainties.  
Aside from electron injector beams like the SuperKEK linac and a potential Linear Collider \cite{Phinney:2007gp}, 
most modern high-intensity electron beams are of the ``continuous wave'' variety, where packets of particles 
$\sim ps$ in duration arrive continuously with $\sim\ns$ spacing, making timing-based rejection of cosmogenic backgrounds rather difficult. 
For example, Jefferson Laboratory's 12 GeV CEBAF beam is of this type \cite{2011JPhCS.312c2014M}. 
Detectors like CDMS-SUF actively shielded against cosmogenic neutrons and vetoed muons, with reduction powers $\sim 5\%$ and $1 \%$ respectively \cite{TAP_THESIS}. A combination of active shielding, neutron vetoes, and a veto on progenitor muons could plausibly reduce the cosmogenic neutron background by a factor of $\sim 10^{3}$ even at a continuous-wave beam.

In the remainder of this section, we will describe backgrounds relevant for neutral current signal detection 
behind an electron beam dump (Sec.~\ref{SubSec: BeamRelatedBackgrounds} and Sec.~\ref{SubSec: BeamUnrelatedBackgrounds}), 
as well as several experimental benchmarks to quantify potential physics reach (Sec.~\ref{SubSec: BenchmarkSetup}); these results
are summarized in Table \ref{tab:backgrounds}.
For concreteness, we will focus on the high intensity electron beam setup appropriate for Jefferson Lab's 12 GeV CEBAF 
and assume a 1m $\times$ 1m $\times$ 1m mineral oil detector placed 20 m behind
the beam dump. Although mineral oil is not necessarily optimal for this experimental setup, it
 is currently used by MiniBooNE  \cite{AguilarArevalo:2010cx} to study quasi-elastic
nucleon recoils from relativistic neutral current processes, so it is a plausible, conservative choice. 
Denser, more responsive materials may be better suited, but their feasibility is not yet established and warrants further study. 
 To achieve optimal sensitivity, such an experiment would require some level of cosmogenic background rejection or shielding given 
 that CEBAF is a ``continuous wave'' beam.  
We will also comment on similar setups at a future Linear Collider or potentially even at the SuperKEK $7 \ \GeV$ injector. 

\subsection{Beam Related Backgrounds}
\label{SubSec: BeamRelatedBackgrounds}

Beam related backgrounds consist of neutrons, muons, and neutrinos that can penetrate the beam dump (or emerge
obliquely, but re-scatter from the atmosphere) and reach 
the detector coincident with beam bunches. The most significant of these are neutrons produced in the dump target --- these rapidly lose energy down to the $\sim 1-5 \ \MeV$ level, at which point they tend to scatter semi-elastically in materials like rock or air. 
These ``fast'' neutrons can reach a nearby downstream detector (usually with a time delay), 
as was observed for example by mQ \cite{slac-mQ,*SLAC-thesis}.
This background does not fake our recoil signal as it is below the recoil energy cuts we apply.

\begin{table}[t] 
\begin{center}
\begin{tabular*}{0.5\textwidth}{@{\extracolsep{\fill}}lcl}
	\hline\hline
	\\[-7pt]
	 $\quad$  {\bf Beam Related}  &  & {\bf Relevance} \\[2pt]
	\hline
	\\[-6pt]
	 ``Fast" and "slow" neutrons  && Remove with $Q$ cut   \\[2pt]
             Stopped  $\mu \to \nu X$ decays & & Remove with $Q$ cut \\[2pt]
             Boosted  $\mu \to \nu X$ decays & &  Rate $\lsim 10^{-7}$ Hz  \\[2pt]
             Boosted  $\pi \to \nu X$ decays & &  Rate $\lsim 10^{-7}$ Hz  \\[2pt]
             	 	\\[-6pt]
	 \hline \hline
	\\[-6pt]
	 $\quad$  {\bf Beam Unrelated}  &  & {\bf Relevance} \\[2pt] 
	 \hline 
	  Radiological (dirt/rock) & & Remove with $Q$ cut   \\[2pt]
	  Cosmic $\mu$ (passing through) & & Rate $\sim 5\cdot 10^{-3}$ Hz, veto	\\
	  Cosmic $\mu$ (stopped decay) &  	&   Timing veto $\sim$ 100 $\mu s$	\\
	  Cosmogenic neutrons  & & Rate $\sim 2 \cdot 10^{-2}$ Hz   \\[2pt] 
	  \hline
	  	\end{tabular*}
\caption{\label{tab:backgrounds} List of dominant backgrounds for the benchmark JLab scenario with $10^{22}$ EOT, a 12 GeV continuous wave beam, and a 1 m$^3$ mineral oil detector situated 20 m downstream at a depth of 10 m.w.e.  For a detailed discussion of beam related and beam unrelated processes see Sec.~\ref{SubSec: BeamRelatedBackgrounds} and Sec.~\ref{SubSec: BeamUnrelatedBackgrounds}
respectively. 
}
\end{center}
\end{table}

For a meter-sized detector positioned $\sim$ 20 m behind the target, muons (and high energy neutrons) produced in a Hall A style beam dump with beam energy of 12 $\GeV$ will range out before reaching the detector, so we need not consider them for 
our benchmark scenario. For alternative setups with either a closer detector or higher beam energy, this 
background may be important, but can still be reduced by actively identifying and rejecting muons, or using
 a deflector magnet to divert penetrating muons away from the detector. 

\medskip
{\bf\noindent    Neutrino Scattering}
\medskip

The main irreducible beam-related background for high energy $Q^2\gsim (140 \ \MeV)^2$ quasi-elastic $\chi$-nucleon 
scattering signals arises from scattering of high-energy neutrinos produced in the beam dump. 
For proton beams, this background is large, typically dominating over other processes.
The situation is much better for electron beams. To gain some intuition, we need to estimate the beam related
 neutrino flux impinging on our detector  and the interaction probability for 
a neutrino with incident energy at or above the 
$\sim 70 \ \MeV$ threshold required to produce a nucleon recoil (i.e. from backscattering) that passes the
 cuts from \ref{Sec: ProductionDetection}.
For each estimate in the following discussion, we
 will round {\it pessimistically} to error on the side of overestimating this background. 
  
 The neutrino-nucleon cross section in this range varies from 
$\sigma_{\nu-N}\approx 10^{-39} (\frac{E_\nu}{100 \ \MeV})^2$ cm$^2$ for neutrino energies  
below a $\GeV$ to $\sigma_{\nu-N}\approx 10^{-38} $ cm$^2$ in the $1-10~\GeV$ range \cite{PDG}.
 The column density of an oil or plastic $1\m$ detector is roughly 
 $85 \ {\rm g}/{\rm cm}^2$, so the integrated nucleon luminosity per incident neutrino is $\sim 5\cdot 10^{25}$ cm$^{-2}$.
Thus, an upper bound on the relevant interaction probability is $\sim 5\cdot 10^{-14}$ to $\sim 5\cdot 10^{-13}$ for neutrinos in the $100 \ \MeV - \GeV$ range respectively.

\medskip
{\bf\noindent Neutrino Production}
\medskip

The largest source of neutrinos with energies near $70 \ \MeV$ is pion electro-production (through the $\Delta$ resonance)
with a cross section of order $\sigma_{eN\rightarrow e+\pi+X}\sim \frac{\alpha}{\pi}\sigma_{\gamma N}$ where 
$\sigma_{\gamma N}\sim 0.6$ mb  
 is the cross-section (per proton) for photon-nucleon inelastic scattering above the pion threshold \cite{PDG}.  
The resulting pions typically stop in several pion interaction lengths before decaying to a neutrino and muon. The muon 
then also stops and decays to two neutrinos and an electron, but these processes can only yield 
neutrinos with energies below $m_\mu/2 \sim 50$ MeV, which is below our benchmark cuts. 
The part of the pion production with boost above $\gamma\sim 2$ that decays before stopping,
which is roughly $3 \%$ (i.e. suppressed by the interaction length compared to lifetime), can give a 
neutrino above $70 \ \MeV$, but this is small compared to the stopped pion component. 

Although the neutrinos that emerge from pion (and muon) decays at rest always carry energy below $70 \ \MeV$, 
 to be very conservative, we assume that every pion yields a neutrino with energy near this kinematic boundary. The 
 stopped pions (and muons) decay isotropically in the beam dump, so we apply an angular acceptance factor of $\sim 2\times 10^{-4}$ 
applicable to the benchmark detector we consider.
The integrated (proton nucleon) luminosity per incident electron on the characteristic stopping distance of three radiation lengths of aluminum 
is $2 \cdot 10^{25}$ cm$^{-2}$. 
The total number of pions produced per incident electron-proton collision is set by $\sigma_{eN\rightarrow e+\pi+X}\sim 2\mu b$, 
so the probability per incident electron is $\sim 4\cdot 10^{-5}$. 
Combining this with the above acceptance factor of $\lesssim2\times 10^{-4}$ and detector interaction probability of $5/4\cdot 10^{-14}$ (for $50 \ \MeV$),
we find a total probability per incident electron of a neutrino scattering event in the detector due to pion production in the beam dump 
of $\sim 1 \times 10^{-22}$. Thus $10^{22}$ electrons on target (EOT) will give a most O(1) neutrinos scattering events, but phase 
space suppression will reduce this considerably, and these events would be below the energy cuts we consider. 

To get above the energy cuts, pions (see below for muons) must carry a boost factor of $\gamma\sim 2$ and decay before stopping. 
This occurs with a suppression of roughly $3\%$ for the in-flight decay, so we expect O(0.03) events, again before 
phase space suppression is taken into account. As the boost factor grows, the pion production cross section falls faster than 
$\propto 1/\gamma^2$, the in-flight decay probability scales as $\propto 1/\gamma$, the acceptance scales as $\propto \gamma^2$, 
and the neutrino scattering probability scales as $\propto \gamma^2$.
Thus, the high energy tail of pion production with $\gamma\sim 10$ can contribute at most O(0.3) neutrinos scattering events in $10^{22}$ EOT,
but again this is an overestimate.  

There is also a high energy component of forward going neutrinos with energy in the $\sim 1 \ \GeV$ range with higher detector interaction probability. This component is dominated by the decay of forward peaked high energy muons produced through QED Bethe-Heitler reactions of the incident electrons. Because the primary energy loss mechanism of a relativistic electron in matter is QED bremsstrahlung, we can compare the Bethe-Heitler muon pair production cross-section with QED bremsstrahlung to estimate this yield.
This ratio is $\sim (m_e/m_{\mu})^2\frac{\alpha}{2\pi}\approx 3 \cdot 10^{-8}$ \cite{Bjorken:2009mm}.
The produced muons carry energy comparable to the incident beam energy ($1 - 10 \ \GeV$ for example) and stop in less than 20 m. 
The probability that these high energy muons decay before stopping is roughly $2000 \ cm / \gamma c \tau_{\mu}\sim 10^{-2}$.
Assuming all of the resulting neutrinos pass through the detector, and combining with an interaction probability of
 $\sim 5\cdot 10^{-13}$, we find a total probability per incident electron of $\sim 1.5\cdot 10^{-22}$, which, again,
  ignores additional acceptance factors. Therefore, this neutrino background is also negligible. 
  These estimates of the neutrino scattering background are compatible with the full simulation results for neutrino backgrounds in the mQ setup \cite{slac-mQ,*SLAC-thesis}, re-scaled for the energy difference and geometric acceptance. 

In summary, beam related backgrounds for quasi-elastic nucleon scattering at $Q^2\gsim (140 \ \MeV)^2$ is negligible, 
in sharp contrast to the case of a proton beam dump, where the flux of neutrinos is orders of magnitude higher. Importantly,
the kinematics of the signal region should avoid the lower energy slow neutron background from the beam. 

\subsection{Beam Unrelated Backgrounds}
\label{SubSec: BeamUnrelatedBackgrounds}

Most beam dump experiments operate at shallow depths, making backgrounds associated with cosmic-ray muons or cosmogenic neutrons in the detector quite substantial; 
 there are also backgrounds from radiological processes in the surrounding rock and dirt, but these are easily removed by the  $Q^2\gsim (140 \ \MeV)^2$ cut on nuclear recoils.
  For concreteness, we estimate backgrounds on a $1\m\times 1\m\times 1\m$ cubic detector situated under an overburden of 10 meters of rock, or roughly 15 m.w.e.  This roughly corresponds to the depth of Jefferson Lab Hall A beam dump \cite{HallA} and, conveniently, to the depth of the CDMS surface facility (SUF) at which muon and neutron fluxes in the energy range of interest have been extensively studied \cite{Chen,TAP_THESIS}.

\medskip
{\bf\noindent Cosmic Muons}
\medskip

The angle-integrated cosmic-ray muon flux at 15 m.w.e. is roughly $50/\m^2/\s$ \cite{TAP_THESIS}.  
At this depth, the muon flux below $\sim 5$ GeV is roughly flat.  The typical muon, with energy $\sim 2 \ \GeV$, has a lifetime $\gamma c\tau \sim 10^4$ m; thus the total rate for muon decays in flight in a $1\m^3$ detector is $\sim 0.005 \Hz$, which can in any case be vetoed with high efficiency.  A more significant electromagnetic background is the stopping and subsequent decay of muons - roughly 10 \% of them - in the detector.  The vast majority of these can be rejected by vetoing on near coincidences with a muon hit near the detector. Vetoing on muon hits in a window as large as 100 $\mu s$ should catch essentially all stopped muon decays, with negligible effect on detector livetime.

\medskip
{\bf\noindent Cosmogenic Neutrons}
\medskip

For the quasi-elastic $\chi$-nucleon collision signals considered here, cosmogenic neutrons are a more directly relevant background.  
Neutrons entering the detector with kinetic energy above $\sim$10 MeV could potentially fake the quasi-elastic $\chi$-nucleon signal process.  
Monte Carlo simulations of neutron flux at SUF \cite{TAP_THESIS} suggest a flux of about $2\times 10^{-2}\m^{-2} \s^{-1}$, with the lower-energy neutrons arising from muon capture in nuclei, and the higher-energy neutrons originating from hadronic showers in cosmic-ray-nucleus interactions.  With no background rejection, this flux over a $2\cdot 10^{7}$ s live time would result in roughly 400,000 nuclear-scattering background events in the experiment's lifetime.   Although a background rate uncorrelated with the beam can be subtracted away, even statistical fluctuations in a  rate this large quickly become the limiting factor for a search, unless the background neutron rate can be greatly reduced.  Some combination of an active neutron veto detector and neutron shielding (and perhaps use of the recoil direction, depending on the detection method) could be used to reject these neutron backgrounds.  Even with these measures, we expect that cosmogenic neutron backgrounds would still be a limiting factor for any experiment of this type using a CW beam.  

A great reduction in background could be readily obtained by using a pulsed beam with duty cycle $10^{-2}-10^{-5}$, if the search is designed to trigger on the beam and therefore accumulates comparable signals in a small fraction of the live time.  This approach is used at accelerator-based neutrino experiments, making beam-unrelated backgrounds rather small for these experiments \cite{MiniBooNE_A,*MiniBooNE_B}.  For example, the MiniBooNE beam consists of bunch trains of order $\sim \mu s$ in duration, spaced by $\sim ms$, with each bunch train containing packets of protons   spaced by $\sim 100$~ns.  Likewise, linear collider beams (or perhaps the SuperKEK injector) are by design both low-duty-factor and high-current, and are therefore well-suited to this purpose \cite{Phinney:2007gp}.  

\subsection{Benchmark Setup}
\label{SubSec: BenchmarkSetup}

\begin{figure}[t]
 \vspace{0.1cm}
  \includegraphics[width=8.6cm]{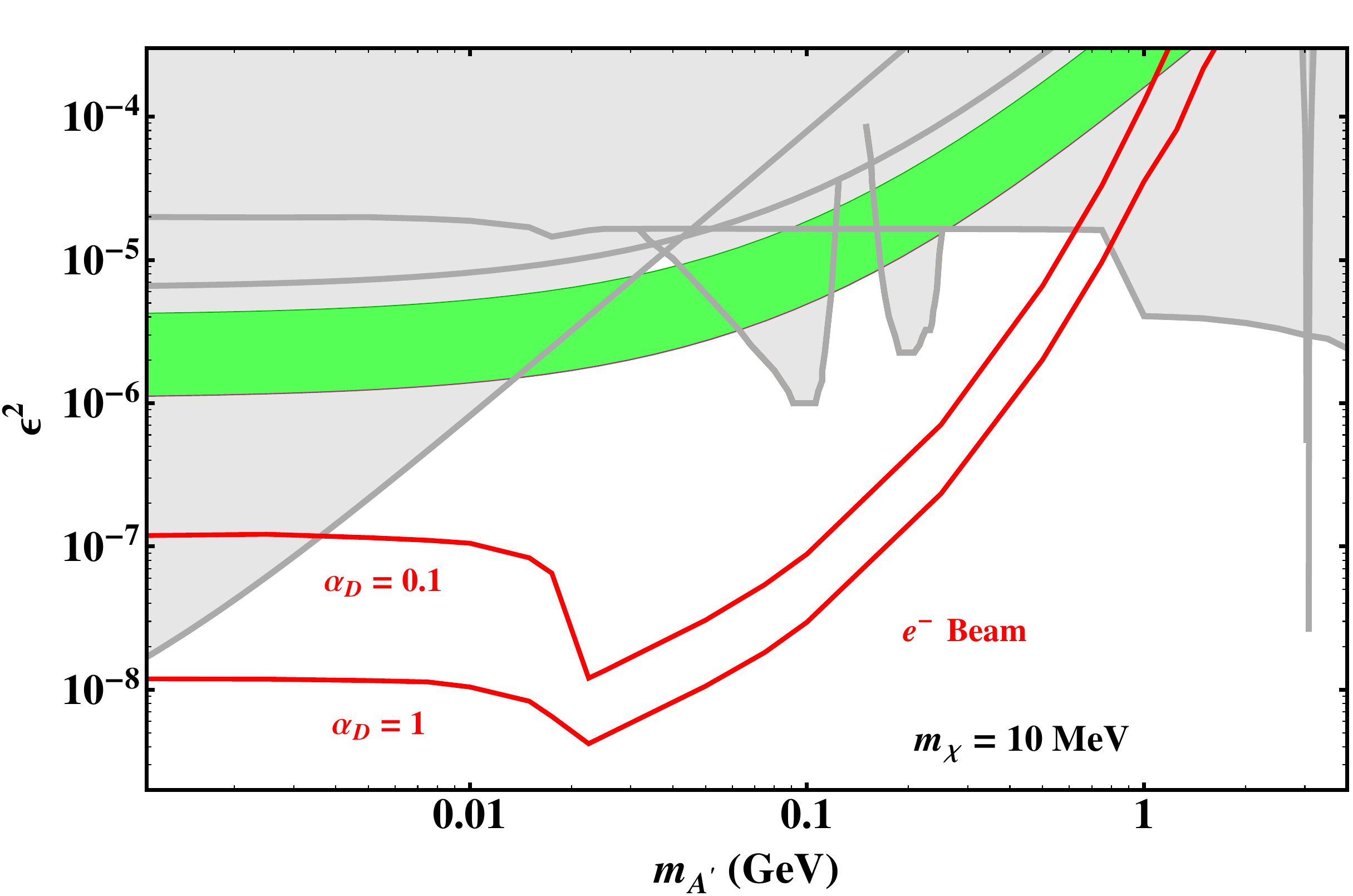} \\
   \includegraphics[width=8.6cm]{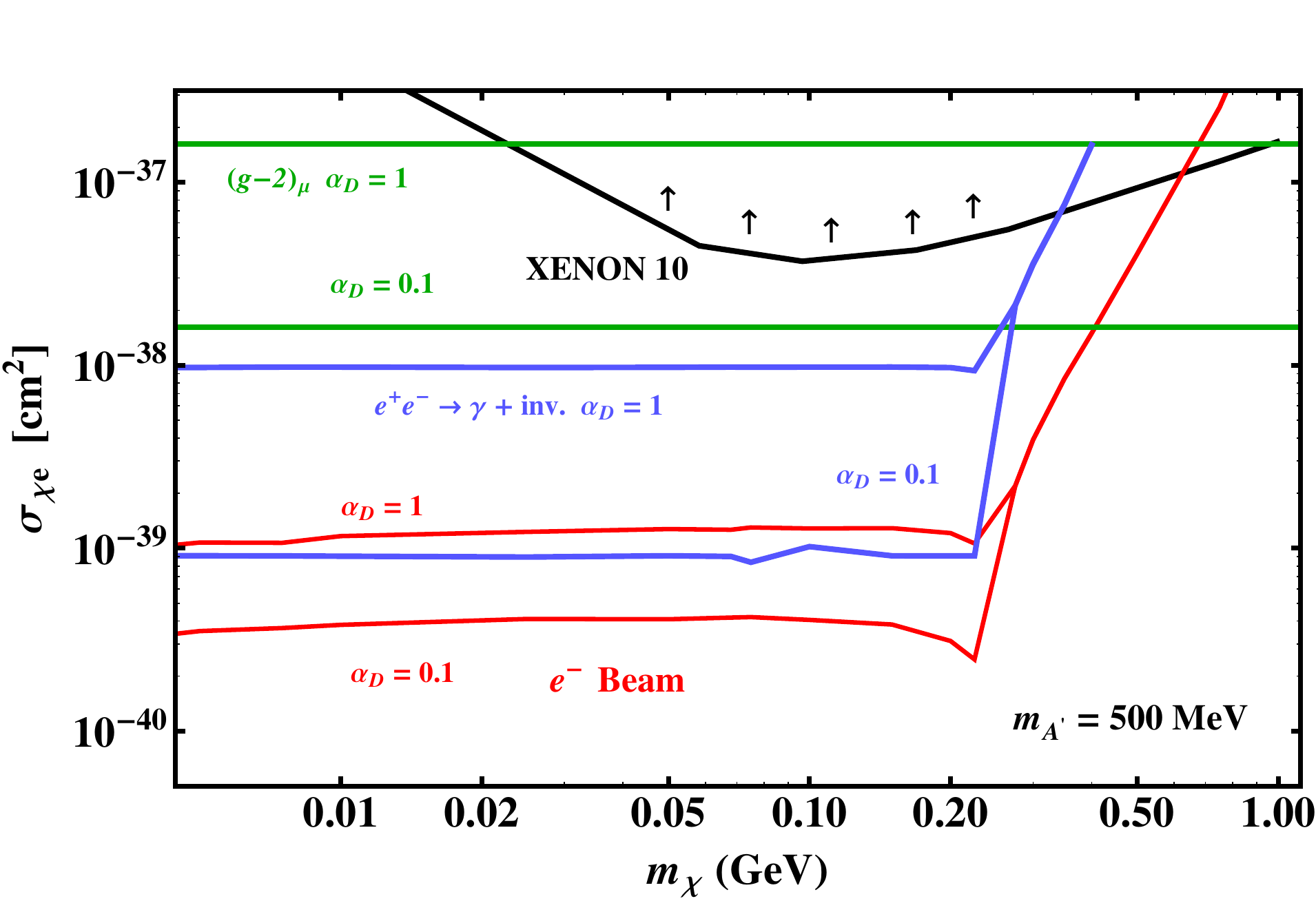}
  \caption{
 Same as Figure \ref{fig:SummaryReach} with overlays of curves for $\alpha_D = 1, 0.1$ The gray region is excluded
 by the same curves shown in \ref{fig:SummaryReach} and described in the caption. 
  }\label{fig:VaryAlpha}
\end{figure}

For concreteness, we have estimated the sensitivity of a small-scale experiment that might be feasible behind the Jefferson Lab Hall A or C beam-dumps, using the 12 GeV CEBAF electron beam.  The basic setup is depicted in Fig. \ref{fig:setup}.  For beams of 12 GeV or lower energy, a detector placed 20 m downstream of the beam dump will have negligible beam-related backgrounds, as this distance significantly exceeds the range of higher energy muons and neutrons in rock, and neutrino rates are vanishingly small.  As was noted earlier, cosmogenic backgrounds are significant for experiments of this type using continuous-wave (CW) beams with high duty cycle.  
High-current running is therefore advantageous to increase the signal yield relative to these backgrounds.  
Jefferson Lab experiments frequently run at beam currents of $50-100 \mu A$; an average current of $80 \mu A$ over $2\cdot 10^{7} \s$ corresponds to a total charge of $10^{22}$ electrons on the beam-dump target (EOT).  Accelerator and installation down-time, experiments using lower beam currents, and lower-luminosity calibration runs would provide additional windows in which to measure the beam-unrelated backgrounds.  Accounting for this duty cycle, $10^{22}$ EOT is a reasonable expectation for an experiment in place for 2 to 3 years.  

With these considerations in mind, we consider a reference geometry with a $1 \m^3$ pure mineral oil detector situated $20 \m$ downstream of a beam dump.  The beam is modeled as a monochromatic $12 \GeV$ electron beam scattering off Aluminum (the main component of the JLab Hall A beam dump \cite{HallA}).  The $\chi$ yield is computed by simulating electron-nucleus scattering at $12 \GeV$ and multiplying by the column number-density of one radiation length of Aluminum, $24 \g / \cm^2$, and by the integrated electron flux of $10^{22}$ EOT.  For the resulting $\chi$ within the geometric acceptance of the detector, their quasi-elastic scattering rate is modeled using the formulas of Appendix \ref{sec:AppendixA}.  Detector efficiency is not included for either signal or background events, but is $O(1)$ for other mineral-oil detectors \cite{MiniBooNE_A,*MiniBooNE_B}.  
We do not explicitly include the effects of beam straggling, but our use of a single radiation length as the effective thickness of a thick target {\it with straggling} 
compensates for this. This is justified in Appendix \ref{appss:production}. We neglect showering effects, but this only {\it underestimates} the total signal yield as the scattering of secondary electrons from the electromagnetic shower produces additional $\chi$. In general, our mock-up of straggling and not including showering are expected to underestimate 
the yields for $m_{A'}$ and/or $m_{\chi} \lesssim 0.1 \MeV$, while we expect good agreement for higher masses. 

In Fig. \ref{fig:SummaryReach} we present our experimental sensitivity to $m_{A^\prime}$ and $m_\chi$ as functions of $\epsilon^2$ and 
 the $\chi$-$e^-$ direct-detection cross section $\sigma_{\chi e} \equiv 16 \pi \epsilon^2 \alpha \alpha_D m_e^2 / m_{A^\prime}^4$ respectively, 
 for three different background rejection scenarios elaborated below.  
The solid, dashed, and dot-dashed red curves mark the parameter space for which our benchmark setup yields 40, $10^3$, and $2 \times 10^4$ signal events
   respectively, which correspond to $2\sigma$ exclusion sensitivity for the three scenarios. The last of these corresponds to a scenario with no background rejection at all.
These curves demonstrate that, even without additional background rejection, our benchmark scenario has interesting physics reach. 
Note the total signal rate is proportional to $\epsilon^4$ and the vertical axes in both plots scale as $\epsilon^2$, so requiring ten times as many signal events
  corresponds to shifting the $\epsilon$ reach upward by a factor of $\sqrt{10}$.  For the same reason, the $5\sigma$ local significance 
  sensitivity of each benchmark scenario would be only a factor of 1.5 higher than the line shown on the plot. 
\begin{itemize}
\item \textbf{Benchmark A} is meant to be representative of a test-stage $1\m^3$ detector with no shielding or veto on cosmogenic neutrons.  Following the estimates above, such a detector would record $N_{bkg} \sim 4\cdot 10^{5}$ cosmogenic neutron events in $2\cdot 10^{7} \s$ of live-time.  In the absence of systematic uncertainties, such an experiment would have exclusion sensitivity to signals at the level of $2\sqrt{N_{bkg}} \sim 1200$ events, but it is more plausible that a parasitic experiment would encounter few-percent systematic uncertainties in this beam-unrelated background rate.  With a $2.5\%$ systematic uncertainty on the neutron flux during beam-on periods, such an experiment would have $2\sigma$ exclusion sensitivity at the level of $20,000$ events per $10^{22}$ EOT.  This sensitivity is denoted by the dotted red lines in Figures 2(a) and (b).  That such a simple experiment has any new sensitivity at all demonstrates the efficacy of this search strategy, but of course the aim of any search program of this type would be to reduce backgrounds significantly more. 
\item \textbf{Benchmark B} demonstrates the benefits of a factor of $1/20$ suppression of the cosmogenic neutron background, with the same $1\m^3$ fiducial volume.  A factor of 20 rejection is only slightly better than was obtained using shielding in the CDMS-SUF detector \cite{TAP_THESIS}, and also appears to be achievable using an active neutron veto.  A $2.5\%$ systematic uncertainty on the neutron background of $20,000$ events would still dominate over statistical uncertainty, allowing $2\sigma$ sensitivity to a 1000-event signal.  This sensitivity curve is denoted by the dashed red lines in Figures 2(a) and (b).  
\item \textbf{Benchmark C} illustrates the sensitivity of a $1\m^3$ fiducial volume detector with aggressive background suppression, at the level of $10^{-3}$ for only 400 background events during live-time, after selection.  In this case, statistical uncertainties dominate for $10^{22}$ EOT, leading to $2\sigma$ exclusion sensitivity at the level of $40$ events.  The resulting sensitivity curve is denoted by the solid red lines in Figures 2(a) and (b).  Clearly, cosmogenic neutron rejection at the $10^{-3}$ level would open up considerable new parameter space. 
\end{itemize}

The orange curve in Fig. \ref{fig:SummaryReach} marks the parameter space for which the same setup yields 10 signal events 
with a first stage pulsed ILC style beam operating at 125 GeV \cite{Phinney:2007gp}. For the same luminosity, the live time for this beam can easily 
be reduced by a factor of $\sim 10^4$, which brings environmental backgrounds down to the $\lsim 100$ event level, so
this experiment can reasonably be sensitive to 10 signal events. 

\begin{figure}[t]
 \vspace{0.1cm}
  \includegraphics[width=8.6cm]{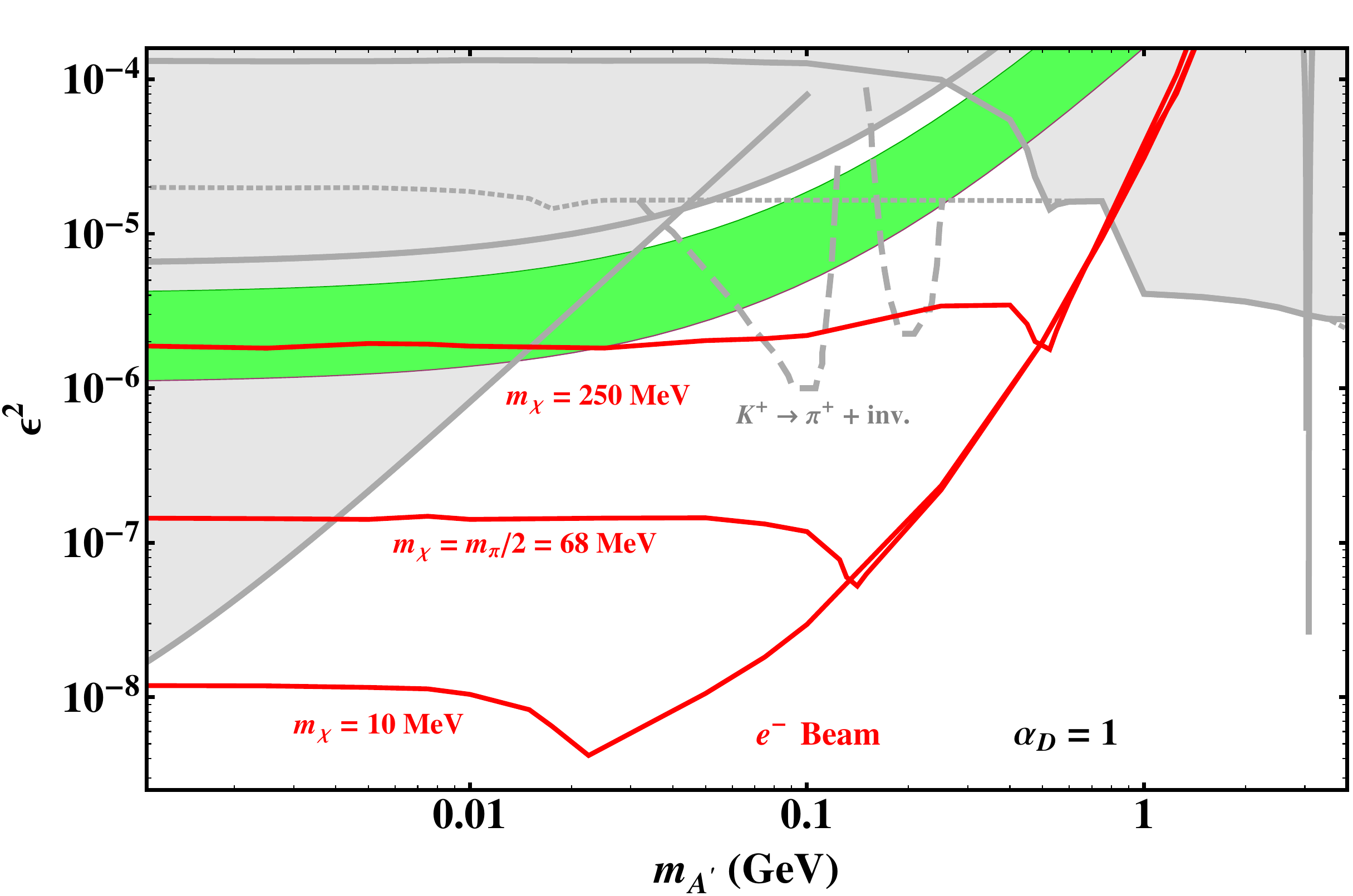} \\ 
    \includegraphics[width=8.6cm]{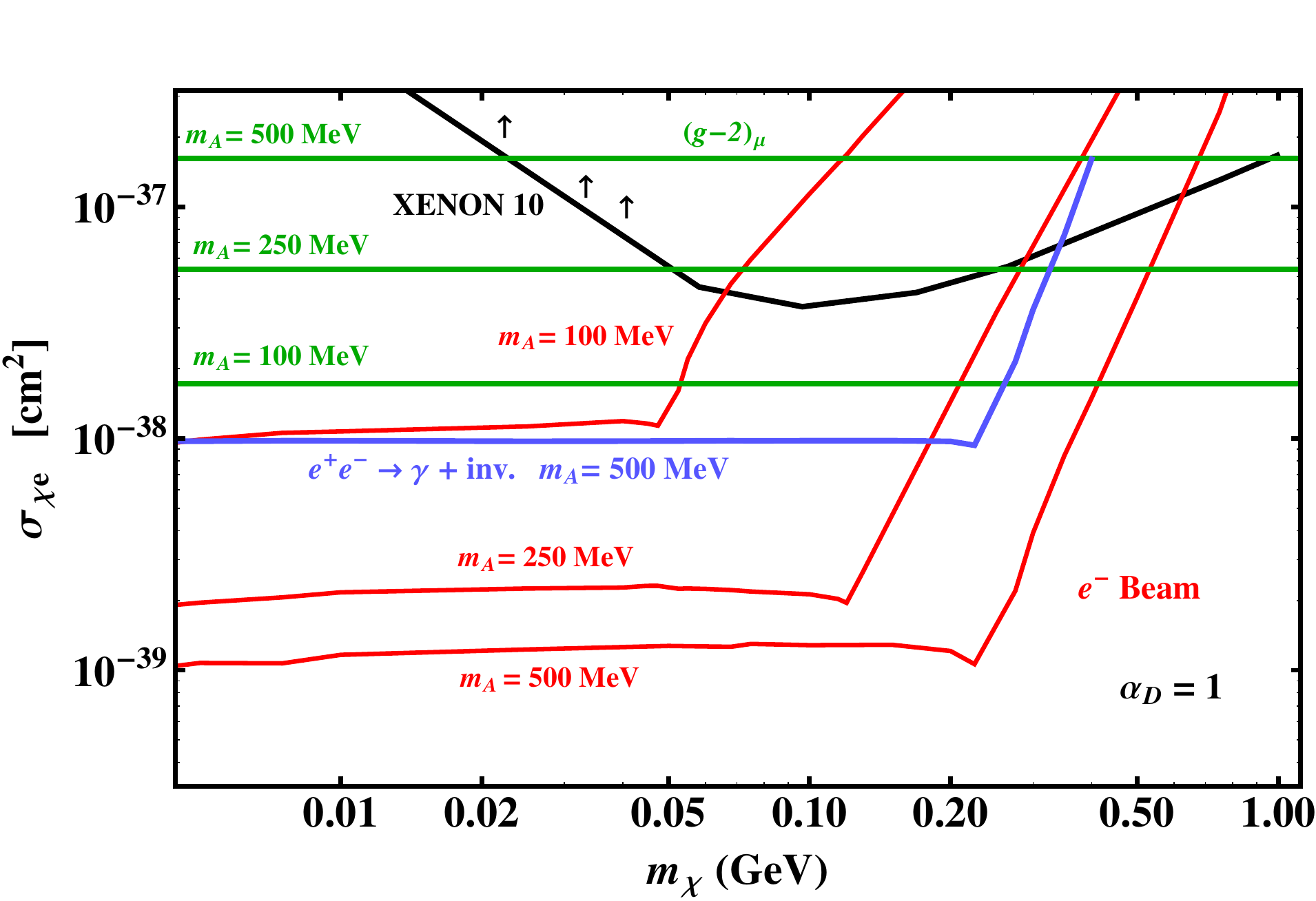}
  \caption{
 Same as Figure \ref{fig:SummaryReach} with overlays of different $m_\chi$ (top) and
 $m_{A^\prime}$ (bottom) parameters.  The $m_\chi = m_\pi /2 = 68$ MeV curve in the top in a) shows 
 our $\epsilon$ reach in the regime where the dominant $\pi^0
 \to \chi \bar\chi$ process is kinematically forbidden at proton beam experiments.  The dotted 
 curve shows the BaBar $e^+e^- \to \gamma +$ invisibles for $m_\chi = 10$ and 68 MeV. This 
 constraint is identical for all $\chi$ produced in on-shell $A^\prime$ decays; for $m_\chi$ 250 MeV this bound shifts 
 upward. Similarly, the kaon decay constraints are now dashed since both regions apply for $m_\chi = 10$ MeV, but 
 neither does at 250 MeV. }\label{fig:VaryMchi-VaryMA}
\end{figure}

\begin{figure}[h]
 \vspace{0.1cm}
  \includegraphics[width=8.6cm]{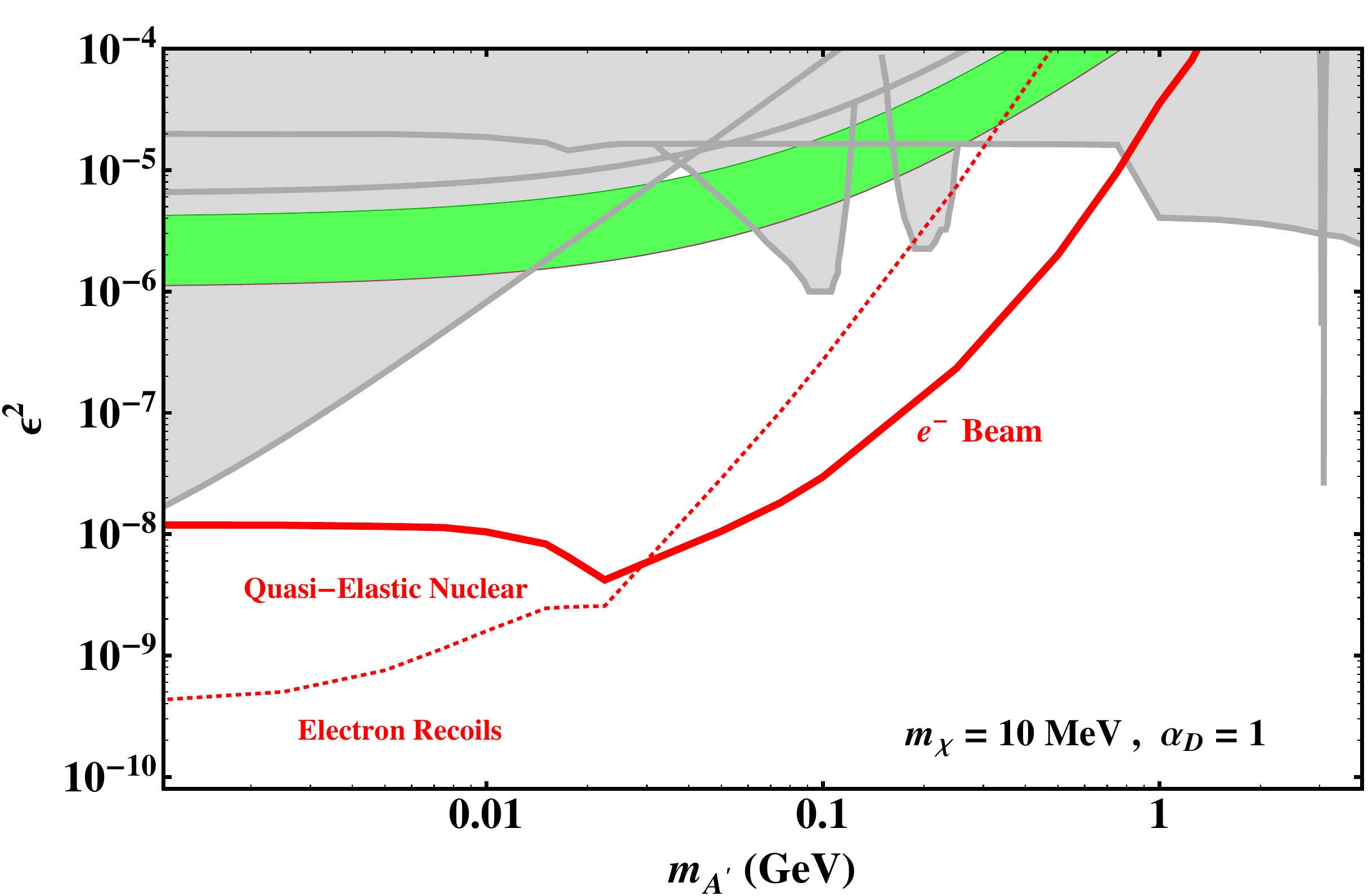} 
  \caption{
 Same as Figure \ref{fig:SummaryReach} (top) with red sensitivity curves for 
 inelastic nuclear (thick, solid) and electron (dotted) signal yields. 
  }\label{fig:AllProcesses}
\end{figure}

For a given experimental setup, the sensitivity depends only on the Lagrangian parameters
in Eq.~(\ref{eq:lagrangian}), which determine $\chi$ production rates, 
scattering probability, and geometric acceptance. 
In every signal process, the $\chi$ scattering rate is proportional to $\epsilon^2\alpha_D$, 
so increasing $\alpha_D$ always yields greater $\epsilon$ sensitivity.  
However, for $m_{A^\prime} < 2 m_\chi$, the $\chi$ production rate is linear in 
$\alpha_D$, whereas the on-shell process  ($A^\prime \to \chi \bar \chi$) is set by
 the $A^\prime$ production rate, which is less sensitive to $\alpha_D$, so the relative
reach in $\epsilon$ can differ substantially in these regimes. 
In Fig. \ref{fig:VaryAlpha} we plot the 
sensitivities for $\alpha_D = 1$ and 0.1 as we vary  $m_{A^\prime}$ with fixed $m_\chi$ (a) 
and vice versa (b) in the benchmark C scenario.  For a fixed signal yield, the off-shell $\epsilon$ reach scales as $1/\sqrt{\alpha_D}$ and
 $\sigma_{\chi e} \propto \epsilon^2 \alpha_D$, so the red curves lines in 
 Fig.~\ref{fig:VaryAlpha}~b) are independent of $\alpha_D$ in this regime. 

 In Figure \ref{fig:VaryMchi-VaryMA} we vary $m_{A^\prime}$ for 
 various choices of  $m_\chi$ (a) and vice-versa (b) in the benchmark C scenario. 
 When $m_{A^\prime} \ll 2 m_\chi$ the production is 
   off-shell and the momenta in all $A^\prime$ propagators is $\sim 2 m_\chi$  
    so the red curves are flat as neither the production nor detection rates depend on $m_{A^\prime}$;
    the sensitivity is set entirely by $m_\chi$.
    For $m_{A^\prime} \approx 2 m_\chi$ the simultaneous resonant enhancement in production and non-relativistic $\chi$ enhancement to acceptance noticeably improve $\epsilon$ sensitivity near threshold. 

\subsection{Experimental Considerations for Future Study}

Although our discussion has emphasized quasi-elastic $\chi$-nucleon scattering in the detector, electron scattering, coherent nuclear scattering, and inelastic collisions that produce $\pi^0$ or $\pi^+$  may also be important signatures if backgrounds for these processes are manageable. 
In Figure \ref{fig:AllProcesses} we show the benchmark C sensitivity for the quasi-elastic and electron-scattering processes. 
In particular, higher energy $\chi$-scattering reactions that produce pions may be significantly easier to see 
above lower energy cosmic neutron backgrounds. While the total signal yield may be lower compared 
to quasi-elastic reactions, cosmic neutron background rates fall rapidly with energy. 
Additionally, higher energy inelastic reactions that produce pions carry more directional information, 
allowing further suppression of cosmic backgrounds.  
Given that background systematics will likely dominate the sensitivity for 
first generation versions of these experiments, using such signals with lower intrinsic background rates 
may be advantageous.  Signals for $\chi$ scattering through $O(\GeV)$ dark photons will induce relatively high-energy nucleon recoils even in quasi-elastic $\chi$-nucleon scattering --- a spectrum quite different from that of cosmogenic neutron backgrounds.  An analysis optimized to distinguish these signals could likely achieve better high-mass sensitivity than what is shown in the figures.

For experiments that are limited by cosmic backgrounds, it may be better to situate 
the detector closer to the dump than what we consider, even though penetrating muons 
can reach the detector and beam related neutrinos may contribute O(1) events. 
Provided muons can be vetoed (or deflected), this would improve acceptance without compromising sensitivity -- after all, 
the cosmic background rates determine sensitivity in this case as beam related backgrounds will continue 
to be negligible {\it by comparison}.  The improvements in sensitivity would be most noticeable in 
the higher $m_{A'}\gsim 100 \ \MeV$ part of the parameter space as compared to our layout. 

Detectors with smaller fiducial volume than our benchmark $1\ \m^3$ should also be considered, as they may be easier to shield from neutrons or surround by a muon veto.  
This may improve sensitivity, even though the yield decreases with detector volume (for high enough $A'$/$\chi$ masses that angular acceptance is relevant) or thickness (for lower $A'$/$\chi$ masses). Background rates will also fall by virtue of the smaller size, and if the shielding/veto reduces backgrounds further then the overall sensitivity of a small detector may exceed that of a larger one.

All of these considerations warrant further investigation. 


\section{Comparison with Neutrino- and B- Factories} \label{Sec:ProtonComparison}

 Proton beams at a variety of existing neutrino experiments  have recently been considered as potential sources of 
  MeV$-$GeV scale dark-sector particles \cite{Batell:2009di,deNiverville:2011it, deNiverville:2012ij,Dharmapalan:2012xp}. In the low beam-energy regime $\lsim 1$ GeV,
the LSND experiment is sensitive to $\chi\bar\chi$ production through rare pion decays $ pp \to \pi^0 + X$, $\pi^0 \to \gamma A^\prime \to \gamma \chi\bar \chi$;
    at intermediate energies near $10$ GeV, MiniBooNE can also probe $\chi\bar\chi$ produced in the analogous $\eta$ decay; and in the high-energy regime
 $\gg10$ GeV, MINOS and T2K can also produce $\chi$ pairs through partonic QCD processes including resonant $q\bar q\to A^\prime\to \chi\bar\chi$
production. These approaches have both advantages and drawbacks relative to the electron-beam scenario outlined in this paper.

 Lower-energy neutrino factories may already be sensitive to new parameter
  space. A forthcoming study of LSND neutral-current signals demonstrates sensitivity to $\epsilon^2 \lsim  10^{-8} -10^{-6}$ for $2m_{\chi} <  m_{\pi^0}$ simply
   comparing predicted signal yields against the background uncertainty \cite{deNiverville:2013}. 
    Sensitivity projections for a dedicated MiniBooNE search reach $\epsilon^2 \lsim 10^{-6}$ for $m_{A^\prime}$ of order 100 MeV or lower \cite{deNiverville:2011it}.
    This effort requires dedicated off-target running \cite{Dharmapalan:2012xp} -- as of the date of this paper, the proposal for new beam time has not been approved.
       
For a given number of charged particles on target, the $\chi$ production yield within acceptance at T2K for $m_{A'}\sim \GeV$ is comparable to that
 of our benchmark electron beam experiment \cite{deNiverville:2012ij}. 
 However, the very neutrino signals that these experiments were designed to measure are now 
  {\it irreducible} backgrounds for other invisibles that undergo neutral current scattering. 
 Unlike MiniBooNE, which runs at $\sim 9$ GeV and can use timing delays of the $\chi$'s to reduce the beam-neutrino background,  T2K operates at $\sim 30$ GeV
 (and the near detectors are closer to the target than MiniBooNE) so that MeV$-$GeV scale particles are too boosted for their arrival times to be distinguished from those of neutrinos.  
 Moreover, the near detectors at T2K (INGRID and ND280) are poorly instrumented for detecting neutral current quasi-elastic scattering.  The on-axis INGRID detector is designed to 
  detect final-state muons from charged-current scattering \cite{Abe:2011xv}. 
  The off-axis ND280 detector may be sensitive to $\epsilon^2 \sim 10^{-5} - 10^{-4}$, but 
   sensitivity is limited by large systematic uncertainties near 30 $\%$ \cite{Short:2013qda} and a dedicated background
   study is necessary to understand the potential reach.  
The MINOS sensitivity to dark matter is similarly hampered by irreducible backgrounds ($\sim 10^6$ neutral current events for $10^{22}$ POT \cite{Adamson:2010wi}) and large systematic uncertainties. 

  An electron-beam production mechanism has several advantages over these approaches. The approach we propose 
  can operate parasitically, on a small scale, and at relatively low cost. Beam neutrinos are negligible in our setup; instead, the
  dominant backgrounds are cosmogenic.  These backgrounds are, in principle, reducible, and can also be measured accurately during beam-off periods
  so that the dominant uncertainties are statistical.  
  Furthermore, $\chi$ electro-production in electron-nucleus collisions 
  is analogous to well-known, perturbative QED processes, so signal production is under good theoretical control.

The complementarity of the beam-dump approach to B-factory searches for invisible final states has already been mentioned in \ref{ssec:collider}.  The two approaches have distinct advantages. B-factories have greater sensitivity to $A'$ and $\chi$ masses $\gtrsim 1\ \GeV$, because the $A'$ production cross-section at B-factories is independent of mass, while the $\chi\bar\chi$ off-shell production rate varies only logarithmically with $m_{\chi}$.  However, the instrumental background from $\gamma\gamma$ events where one photon leaves no signal in the detector --- which is kinematically indistinguishable from sub-GeV-mass $A'\rightarrow invisible$ signals --- poses a serious challenge to searches for sub-GeV $A'$.  This is, of course, the mass range where the fixed-target approach comes into its own, with limited form-factor suppression and reasonably high acceptance for a small detector.  A second difference that bears mentioning is that the $B$-factory on-shell $A'$ rate is independent of $\alpha_D$.  The corresponding fixed-target signal, in contrast, is enhanced at large $\alpha_D$ and suppressed for small $\alpha_D$ because of the $\alpha_D^2$ scaling of the $\chi$ detection probability.  


\section{Conclusions and Discussion} \label{Sec:Conclusion}
Long lived, weakly-coupled particles in the MeV-to-few-GeV range
 are viable and generic dark-matter candidates in extensions of the Standard Model by a new ``dark sector.''  Yet such states remain difficult to probe experimentally.
In this paper we have examined existing constraints on these particles, 
and shown that a fixed-target experiment searching for their scattering in a $\sim 1\m^3$ (or smaller) detector volume downstream of an electron beam-dump can improve sensitivity in this mass range by several orders of magnitude. The proposed experimental setup uses standard technology at currently operating beams and can be run parasitically, without disrupting ongoing physics programs.  
The search strategy is systematically improvable by rejection or shielding of cosmogenic neutron backgrounds, with negligible beam-related backgrounds.  
 
Currently, the strongest constraints on dark states come from a wide range of sources, depending on the mass and interactions of the particle mediating their interactions with the SM. For simplicity, our analysis has concentrated on the benchmark scenario of a kinetically mixed $A^\prime$ that couples to both dark and electromagnetic currents. The strongest constraint 
on $\epsilon$ for $A^\prime$ masses below tens of MeV is from by 
measurements of electron $(g-2)$. For masses above 10 GeV, the strongest bounds are from direct
 detection and collider mono-jet searches. Constraints in the intermediate regime, between a few MeV and a 
 few GeV, are considerably weaker. We have shown that hitherto overlooked B-factory searches in $\gamma+{\rm invisible}$ 
 final states currently set the strongest limits for much of this mass range.  While these searches can be significantly extended at high-luminosity experiments like Belle II, the possible gains for $A'$ masses below $\sim 1$ GeV will be modest because these searches face a large instrumental background from $\gamma\gamma$ events that fake the $A'$ signal.
 
Important regions of this parameter range remain unexplored and call for new search strategies. 
 If dark sector fields couple to leptonic currents, electron-nuclear collisions in a beam dump generate an energetic and rather collimated beam of 
 $\chi$ particles.  For a $\sim 10$ GeV electron beam, muons and energetic hadrons range out 10--20 meters from the dump, and beam-related neutrino backgrounds at these distances are also negligible.  A $\m^3$-scale (or smaller) scintillator detector can therefore be used to detect the neutral-current-like scattering of the $\chi$'s off target nucleons, nuclei, or electrons, with a smooth exchange momentum spectrum dominated near $Q^2 \sim m_{A'}^2$.  This search strategy is quite complementary to $B$-factory searches, which have excellent sensitivity above a GeV, but significant background limitations below a GeV where the beam-dump approach is at its strongest.   

For concreteness, we have illustrated the sensitivity of a mineral-oil detector sensitive to quasi-elastic scattering off nucleons, for which detailed background estimates at the relevant depths exist.  Such an experiment's sensitivity depends strongly on its ability to reject cosmogenic neutron backgrounds, for example using a combination of beam-timing (particularly at pulsed-beam facilities), muon vetos, an active neutron veto, and/or neutron shielding. Optimizing these strategies may involve experimental variations on the baseline detection scenario considered here.  The effects on sensitivity of varying cuts, detector materials and detection techniques, and geometric layout have
 not been considered in detail, and warrant further optimization. Furthermore, sensitivity could improve dramatically by considering electron, coherent nuclear, 
 and inelastic pion signals. Additional gains may also be achieved if directional information (potentially exploiting tracks from charged pions),
  pulse shape discrimination, and ionization yields are used  to distinguish signal from background. We leave detailed studies of experimental variations for future work.

The parameter space that these experiments can explore overlaps with two regions relevant to long-standing anomalies in data.  In particular, such a search can decisively test whether an \emph{invisibly decaying} $A'$ resolves the persistent muon $g-2$ anomaly.  Indeed, even a high-background test version of our benchmark setup will probe most of this parameter range ($\epsilon^2 \sim 10^{-6}-10^{-5}$ for $A'$ masses of 10--100 MeV). These searches can also cast light on the few-MeV dark matter motivated by the INTEGRAL 511 keV excess from the inner Galaxy \cite{Knodlseder:2005yq} and few-GeV dark matter motivated by numerous direct-detection anomalies \cite{Aalseth:2012if, Bernabei:2010mq, Agnese:2013rvf, Ahmed:2010wy,Angloher:2011uu}. 
More broadly, these experiments are sensitive to any sub-GeV dark matter or other light (meta-)stable particle whose interactions with SM matter involve a comparably light mediator.

The electron beam energy has important effects on the sensitivity of an experiment of this form.  Lower-energy beams are less effective for larger $\chi$/$A'$ masses, both because the $\chi$ beam is less collimated and because nuclear coherence in production degrades at lower $A'$ masses for lower beam energies.  This may be partially compensated by the ability to place the detector closer to the beam-dump, raising the possibility of sensitive searches in the lower part of this mass range using either the Mainzer Mikrotron (MAMI) or the ELSA electron accelerator.  Higher-energy beams, by contrast, increase the collimation of $\chi$ production, so that the high-mass acceptance scales as $E^2$, while the threshold mass for loss of coherence in production scales as $\sqrt{E}$.  Of course, the range of charged particles also increases with energy; for sufficiently high energies, a magnet behind the dump to deflect secondary muons may be advantageous.  

Time structure is also an important consideration: pulsed beams accumulate a given charge deposition in significantly less live-time, offering an immediate suppression of the dominant cosmogenic neutron backgrounds above what can be obtained by shielding or vetos while the beam-off windows permit a very precise measurement of this background.  These considerations suggest that the pulsed beams from the Super-KEKB linac injector could be used to achieve a much lower-background version of this search, albeit probably not parasitically.  In addition, a detector situated downstream of an ILC beam dump could exploit its high energy and pulsed beam structure.  The impact of this energy gain on an ILC beam dump experiment's high-mass sensitivity can be seen in the figures.

Our approach has distinct advantages over dark matter searches at neutrino factories. In addition to parasitic running,
which is unavailable at relevant proton beams, electron beam dumps allow smaller detectors close to the dump to achieve high acceptance. 
Neutrino factories take advantage of a pulsed beam structure, which allows
for effective rejection of environmental backgrounds, but dark matter searches using their proton beams 
are dominated by large backgrounds from beam neutrinos
with $\sim 10 - 30 \%$ systematic errors. In contrast, existing 
continuous wave electron beams feature negligible neutrino production, but the lack of timing structure 
complicates cosmic background rejection. Nonetheless, whereas beam neutrinos are a source of irreducible
background, cosmogenic backgrounds can be shielded and measured during beam-off periods. Thus, our
setup can be systematically improved by optimizing the detector shielding, composition, geometry,
 and downstream placement. 

The sensitivity of this type of experiment is particularly exciting when viewed
in the context of other ongoing and planned searches for kinetically
mixed gauge bosons in visible decay modes (either promptly into Standard Model
fermions or into promptly-decaying dark sector states).  Planned
fixed-target searches and analyses of B-factory data will probe much
of the parameter space with $\epsilon \gtrsim 10^{-4}$, provided that
the dark photon decays visibly.  The combined results from all three
search strategies --- pair invariant mass and vertexing searches for
$A'\rightarrow f\bar f$, inclusive multi-lepton searches for $A'$
decays into dark sector states that in turn decay to Standard Model
particles (see e.g. \cite{Andreas:2012mt}), and the recoil-based searches
discussed here and in \cite{Batell:2009di,deNiverville:2011it,deNiverville:2012ij,deNiverville:2013} for (meta-)stable dark sector states ---
will \emph{decisively} test the possibility of new MeV--GeV $U(1)$
gauge bosons with photon kinetic mixing $\epsilon \gtrsim 10^{-4}$.

The simple model \eqref{eq:lagrangian} that we have used as a benchmark throughout this paper captures most of the ingredients relevant to fermionic $\chi$ and vector or (with small adjustments) scalar mediators through which they couple to the Standard Model.  However, it is important to more carefully quantify the sensitivity to other models in future work.  The sensitivity to scalar $\chi$ is qualitatively similar, but weakened by factors of few (depending on $A'$ and $\chi$ masses) because the scalar production matrix element is $p$-wave suppressed and peaked \emph{away} from the forward direction, but qualitatively similar.  Another model deserving of study is that where the $\chi$ is only a pseudo-Dirac fermion or split complex scalar, with two mass eigenstates $\chi$ and $\chi^*$ ($m_{\chi^*} > m_{\chi}$). Such a splitting is generic in models with a broken gauge symmetry, and is also characteristic of a dark-Higgs sector with multiple Higgs fields.
 If the $A'$ interacts mainly through an $A'\chi\chi^*$ vertex, then scattering in the detector will be dominated by up-scattering $\chi N\rightarrow \chi^* N$ and/or down-scattering $\chi^* N\rightarrow \chi N$.  While the qualitative signals and yields are rather similar, distinctive features of the nucleon recoil kinematics deserve a close examination.  

\section*{Acknowledgments}
\medskip
We thank Bertrand Echenard and Yury Kolomensky for very helpful exchanges
 regarding the BaBar result \cite{Aubert:2008as}, Maxim Pospelov and Miriam Diamond 
 for numerous helpful conversations, and Denis Perevalov for helpful explanations regarding MiniBooNE analyses. 
 We also thank the PI staff for support. 
This research was supported in part by Perimeter Institute for Theoretical Physics. Research at Perimeter Institute is supported by the Government of Canada through Industry Canada and by the Province of Ontario through the Ministry of Research and Innovation.

\appendix
\section{Signal Calculation}\label{sec:AppendixA}
\label{sec:xsec}

In this appendix we present in detail the method and formulas used to calculate of $\chi$-scattering signal, including both the full model of $\chi$ production in the beam-dump and the model used to calculate nucleon quasi-elastic and electron-recoil signal yields in the detector.  The notation used is as in Section \ref{Sec: ProductionDetection}.

To compute our signal yields for the fixed-target setup we generate a population of $e^- Z \to e^- Z \chi \bar \chi$ scattering events (where $Z$ is a target Aluminum nucleus) using a version of \texttt{MadGraph 4} \cite{Alwall:2007st} modified by one of the authors and R. Essig to simulate fixed-target collisions.  The relevant modifications to \texttt{MadGraph} are: (1) inclusion of initial-state particle masses, (2) a new-physics model including a massive $A'$ gauge boson coupled to electrons with coupling $e\epsilon$, and (3) introduction of a momentum-dependent form factor for photon-nucleus interactions.
The nuclear-elastic and nucleon quasi-elastic form factors used in the simulation is given in Section \ref{appss:production}, which also justifies the approximation of neglecting electron-beam straggling and using only one radiation-length of the target.  For completeness, the Weizsacker-Williams effective photon flux $\alpha\Phi/\pi$ introduced in Section \ref{ssec:production}, defined in terms of the same form factors, is presented in this appendix as well, but the simulation used to generate plots does not use the Weizsacker-Williams approximation.

The \texttt{MadGraph} simulation yields both an inclusive cross-section for $\chi\bar\chi$ production in the beam dump and a Monte Carlo sample of $\chi$ kinematics with the physically correct differential distribution.  The rate of $\chi$ interactions in the detector is determined by selecting the $\chi$'s within detector acceptance and multiplying their interaction cross-sections (for either $\chi$-nucleon quasi-elastic scattering or $\chi$-electron elastic scattering) by the column number density of nucleons and electrons along their path through the detector.  In the case of $\chi$-nucleon scattering, proton and neutron scattering are modeled separately.  The cross-section formulas used in simulation for the $\chi$-nucleon and $\chi$-electron processes, respectively, are given in Sections \ref{appss:nrecoil} and \ref{appss:erecoil}, with a minimum recoil energy requirement of $10 \MeV$.

In a realistic experiment there may also be additional detection efficiencies in addition to angular acceptance and minimum target-recoil momentum. 
However, even for a large ($\sim$ 1000 m$^3$) mineral oil \cite{AguilarArevalo:2010cx} detector
 these efficiencies are $\sim 0.5$, so we expect a smaller, lower background experiment to be more sensitive, so for our 
numerical studies we have set this additional efficiency to unity.

A similar \texttt{MadGraph} model is used to estimate  $e^+ e^-\to\gamma\chi\bar\chi$ signal yields for the BaBar $\gamma+invisible$ search.  The resulting yields agree quite well with the analytic formulas in the text.  For the on-shell $A'$ signals the full \texttt{MadGraph} cross-section within geometric acceptance is used to compute yields; for off-shell signals only the yield with $m_{\chi\bar\chi} < 1 \ \GeV$ is used to compute the yield that is compared to BaBar limits.

\subsection{Model of $\chi$ Production in Beam Dump}\label{appss:production}
Here we give a brief description of the form factors used both in the full $\chi$-production Monte Carlo and in Section \ref{ssec:production}. For details on its validation, see \cite{Bjorken:2009mm}.

In all of the processes of interest, we can focus on electric form factors for either coherent or incoherent scattering off the nucleus.  
For most energies in question, $G_2(t)$ is dominated by an elastic component 
\be
G_{2,el}(t)= \left(\f{a^2 t}{1+a^2 t} \right)^2
\left(\f{1}{1+t/d} \right)^2 Z^2,
\ee
where the first term parametrizes electron screening (the elastic atomic form factor) 
with $a=111\,Z^{-1/3}/m_e$, and the second finite nuclear size (the elastic nuclear form 
factor) with $d=0.164 \mbox{ GeV}^2 A^{-2/3}$.  
We have multiplied together the simple parametrizations used for each in \cite{Kim:1973he}.  
The logarithm from integrating \eqref{ChiExp} is large for 
$t_{min} < d$, which is true for most of the range of interest.  
However, for heavy $A'$, the elastic contribution is suppressed and is 
comparable to a quasi-elastic term,
\be
G_{2, in}(t)= \left(\f{a'^2 t}{1+a'^2 t} \right)^2 \left(\f{1+\f{t}{4
    m_p^2} (\mu_p^2-1)}{(1+\f{t}{0.71\,{\rm GeV}^2})^4} \right)^2 Z,
\ee
where the first term parametrizes the inelastic atomic form factor and the second 
the nucleon quasi-elastic form factor, and where $a'=773 \,Z^{-2/3}/m_e$, $m_p$ is the proton 
mass, and $\mu_p=2.79$ \cite{Kim:1973he}.  
This expression is valid when $t/4m_p^2$ is small, which is the case for $m_{A'}$ 
in the range of interest in this paper.  At large $t$ the form factors will deviate from these simple parameterizations but can be measured from data. 

The effective photon flux used in the Weizsacker-Williams treatment of $\chi$ production in Sec.~\ref{ssec:production} follows directly from these form factors, with dependence on the $A'$ mass, target nucleus, and beam energy.  The effective photon flux $\chi$ is obtained as in \cite{Kim:1973he,Tsai:1973py} by integrating electromagnetic form-factors over allowed photon virtualities:
 
For a general electric form factor $G_2(t)$ (which we take to be the sum of $G_{2,el}$ and $G_{2,in}$ defined above),
\be
\Phi \equiv \int_{t_{min}}^{t_{max}}  dt \f{t-t_{min}}{t^2} G_2(t) \label{ChiExp}
\ee
(the other form factor, $G_1(t)$, contributes only a negligible amount in 
all cases of interest).   
For most $A'$ masses of interest, the integral in \eqref{ChiExp} receives equal contributions at all $t$ below the inverse nuclear size, and so is logarithmically sensitive to $t_{min}=(m_{A'}^2/2E_0)^2$; typically, sensitivity to $t_{max}=m_{A'}^2$ is subdominant because, for large $m_{A'}$ where the  logarithm becomes small, it is effectively cut off below $t_{max}$ by the large-$t$ suppression of $G_{2}$.  
We note also that for ease of simulation, the kinematics of $\chi$ production is implemented in \texttt{MadGraph} as though the entire nucleus is recoiling, even for quasi-elastic processes. Since the energy transfer to the nucleon is typically much smaller than the energy of the $A'$ or $\chi\bar\chi$ pairs, this effect is not very important.

In finding the total number of $\chi$'s produced, we neglect showering in the target; showering would increase $\chi$ production somewhat. Another effect that is not implemented in our Monte Carlo is the energy loss of the electron beam as it traverses the target (straggling).  To account approximately for the effect of straggling, we compute yield from an effective target thickness of only one radiation length, $T_{\text{eft}}$ of 1, even though the target is in fact much longer. This can be justified as follows. 
Given an incident monochromatic electron beam of energy $E_0$, the beam energy distribution after passing through $s$ radiation lengths of the target is given approximately by \cite{Tsai:1966js}
\be
I(E',E_0,s) &\approx& \frac{1}{E_0}y^{bs-1}bs,
\ee
where $b=4/3$, and $y=\frac{E_0-E'}{E_0}$.
For the small angular size $\theta_D$ of the proposed setup, the angular acceptance scales (for large $A'$ or $\chi$ masses relative to $E_0\theta_D$) as $\propto E'^2$. The cross-section for $\chi\bar\chi$ production varies much more slowly with beam energy, and this variation can be neglected to a good approximation.  
The total $A'$ yield in the detector may then be estimated by treating the beam as monochromatic of energy $E_0$ over a thickness $T_{\text{eff}}$, where $T_{\text{eff}}$ is the integral of $I(E',E_0,s)$ weighted by the ratio of acceptance for electron energy $E'$ vs. $E_0$:
\be
\nonumber
T_{\text{eff}} &=& \int ds \int dE' (E'/E_0)^2 I(E',E_0,s) = \frac{3}{2}\log{2}\approx 1
\ee
The formula above neglects energy-dependence of the cross-section, which cuts off the $E'$ integral at some positive $E_{min}$, but for the parameter space of interest the integral is still close to 1.  
Thus for large $A'$ or $\chi$ masses (where the detector acceptance of \eqref{acceptance} is dominated by the $E^2$ term), the $T_{\text{eff}}=1$ approximation accounts, to a good approximation, for electron-beam straggling through a long target.  For $A'$ or $\chi$ masses small enough that the detector acceptance is $O(1)$, the $T_{\text{eff}}=1$ approximation under-estimates the $\chi$ yield.

\subsection{Nucleon Recoils} \label{appss:nrecoil}
This section describes the model of $\chi$-nucleon quasi-elastic scattering used in the Monte Carlo, based on the results of \cite{deNiverville:2011it}.
Equation \eqref{eq:nucrecoilSimple} is a simplification of these results in the limit of large $\chi$ energy, $E \gg  m_{N},m_{\chi},\sqrt{Q^2}$.
\be \label{eq:nuclear-recoils}
\frac{d\sigma_{_{\chi N}}}{dE_\chi} \!\!&=&\!\! 4\pi \epsilon^2\alpha\alpha^\prime \frac{A(E,E_f) F^2_{1,N}(E_f) + \frac{1}{4} B(E, E_f) F^2_{2,N}(E_f)}{  (m_A^2 + 2m_N E_f   )^2 (E^2-m_\chi^2)    }, \!\!\!\!\!\!\!\! \nonumber \\
\ee
where $E$ ($E_\chi$) is the energy of the incoming (scattered) $\chi$, $E_f$ is the nucleon recoil energy, and 
\be
\!\!\!\!\!A(E,E_f) \!\!&=& \!\!2 m_N E (E-E_f) - m_\chi^2 E_f  ~~~,~~  \\
\!\!\!\!\!B(E,E_f ) \!\!&=&  \!\! - E_f  [    (2E- E_f)^2    + 2m_N E_f - 4m_\chi^2         ]~.~~~ 
\ee
The monopole and dipole form factors are 
\be
F_{1,N}(E_f) &=& \frac{q_N}{(1+2 E_f/m_N)^2} ~~, \\ 
F_{2,N}(E_f)  &=&  \frac{\kappa_N}{(1+ 2 E_f/m_N)^2}~~,
\ee
where $q_p=1, q_n = 0$ and $\kappa_p = 1.79$ and $\kappa_n = -1.9$. For a monoatomic detector material, the nucleon-weighted
differential cross section is
\be
\frac{d\bar \sigma_{\chi N}}{dE_f} = \frac{Z}{A} C_p(E_f)\frac{d\sigma_{_{\chi p}}}{dE_f} + \frac{A-Z}{A} C_n(E_f)\frac{d\sigma_{_{\chi n}}}{dE_f}~,~ \nonumber \\
\!\!\!\!\!\!\!\!\!\!\!\!\!\!\!\!\!\!\!\!\!\!\!\!\!\!\!\!\!\!\!\!\!\!\!\!\!\!\!\!
\ee
where $C_{p,n}(E_f)$ are the efficiencies for detecting proton and neutron recoils and $A$ is the detector material's
atomic mass number. For a typical carbon based detector, 
$C_{p,n}(E_f) \approx 1$ over the $E_f \sim  50 - 500$ MeV range \cite{AguilarArevalo:2010cx}. Swapping 
$E_\chi$ in favor of $E_f$ to obtain the differential rate $d\bar \sigma / dE_f$, the kinematically averaged cross section is 
\be
\langle \sigma_{\chi N}\rangle = 
 \int_0^{E_0}dE  \,  \frac{ dF_\chi}{dE}   \int^{\infty}_{E_{f,0}}\!\! dE_f   \frac{d\bar\sigma_{ \chi}}{dE_f} ~,~
\ee
where $E_{f,0}$ is the minimum cut on recoil energies. 

In a monoatomic detector with nucleon density $n_N$, the  total number of nuclear recoils is 
\be
N_{\rm rec}^N = 2 N_\chi P_{\chi N} =    2 ( N_e  X_0  n_Z     \sigma_{ \chi \chi} )(    L_{d} n_N   \langle \sigma_{\chi N} \rangle)~,~~~~
\ee
where $P_{\chi N}$ is the probability for $\chi$ to scatter off a nucleus.

\subsection{Electron Recoils} \label{appss:erecoil}
This section provides the full formula for $\chi$-electron recoil, of which a simplified version appeared in \ref{eq:erecoil}
In the limit where both $\chi$ and the target electron are relativistic in the CM frame and 
up to corrections of ${\cal O}(m_e^2)$,
 the recoil profile for $\chi e \to \chi e$ scattering in the lab frame is 
\be
 \frac{d\sigma_{\chi e}}{dE_{f}} = 
  4 \pi \epsilon^{2} \alpha \alpha^\prime m_e 
 \frac{  
 4 m_e m_\chi^2 E_f + [m_\chi^2 + m_e (E -E_f) ]^2
 }{    (m_A^2  +    2m_e E_{f} )^2 ( m_{\chi}^{2} + 2  m_{e} E)^2  }  ~, \nonumber \\
\ee
where $E$ is the incoming $\chi$ energy and $E_f$ is the electron recoil energy.
Convolving this result with the kinematics of production and the cut efficiency $\epsilon_e$ for electron recoil detection, the 
kinematically averaged recoil cross section for a single $\chi e$ scattering event is 
\be
\langle \sigma_{\chi e} \rangle =  \int_0^{E_0}dE  \,  \frac{ dF_\chi}{dE}   \int_{E_1}^{E_2}\!\! dE_f   \frac{d\sigma_{\chi e}}{dE_f} ~~,
\ee
where $E_{1,2}$ define the electron recoil cuts and $dF_\chi/dE$ is the normalized energy distribution of $\chi$ particles inside the solid angle from the target to the detector
\be
\frac{dF_{\chi}}{dE} \equiv \frac{1}{N_\chi} \int^{\Omega_c}_0 \!\!\! d\Omega \frac{dN_\chi}{ d\Omega \,dE }~~. ~~
\ee
For a detector of length $L_d$ and electron density $n_e$, the number of electron recoils  per incident  $\chi$ is 
\be
P_{\chi e} =  n_e L_d  \langle \sigma_{\chi e} \rangle ~~~,~~
\ee
For a target material with atomic number $Z$, target density $n_T$, and radiation length $X_0$, the number of $\chi$ 
particles produced for $N_e$ EOT is
\be 
N_\chi = N_e X_0 n_T \sigma_{\chi \bar\chi}
\ee
where $\sigma_{\chi \bar \chi}$ is the total $\chi\bar \chi$ pair production cross section in electron-nucleus collisions. 
Thus, the total number of electron recoil events is 
\be
N_{\rm rec}^e = 2 N_\chi P_{\chi N} =    2 ( N_e  X_0  n_Z     \sigma_{ \chi \chi} )(  L_{d} n_e  \langle \sigma_{\chi e} \rangle)
\ee
where the factor of 2 takes into account $\chi$ and $\bar \chi$ pair production.

\bibliographystyle{apsrevM}
\bibliography{DarkDump}
\end{document}